\documentclass[sigplan, 10pt, screen, pbalance]{acmart}

\usepackage{xspace}
\usepackage{booktabs}
\usepackage{enumitem}
\usepackage[nameinlink]{cleveref}
\usepackage{acmart-taps}

\crefname{figure}{fig.}{figs.}
\Crefname{figure}{Fig.}{Figs.}
\crefformat{section}{\S#2#1#3}
\crefmultiformat{section}{\S\S#2#1#3}{ and~#2#1#3}{, #2#1#3}{, and~#2#1#3}

\newcommand{\sys}{\textsc{ELDR}\xspace}

\copyrightyear{2026}
\acmYear{2026}
\setcopyright{cc}
\setcctype{by}
\acmConference[SIGOPS ATC 2026]{The 2026 ACM SIGOPS Annual Technical Conference}{November 15--18, 2026}{Hong Kong, China}
\acmBooktitle{The 2026 ACM SIGOPS Annual Technical Conference (SIGOPS ATC 2026), November 15--18, 2026, Hong Kong, China}
\acmDOI{10.1145/3828523.3857077}
\acmISBN{979-8-4007-2818-1/2026/11}

\acmBadgeR[https://sysartifacts.github.io/atc2026/results]{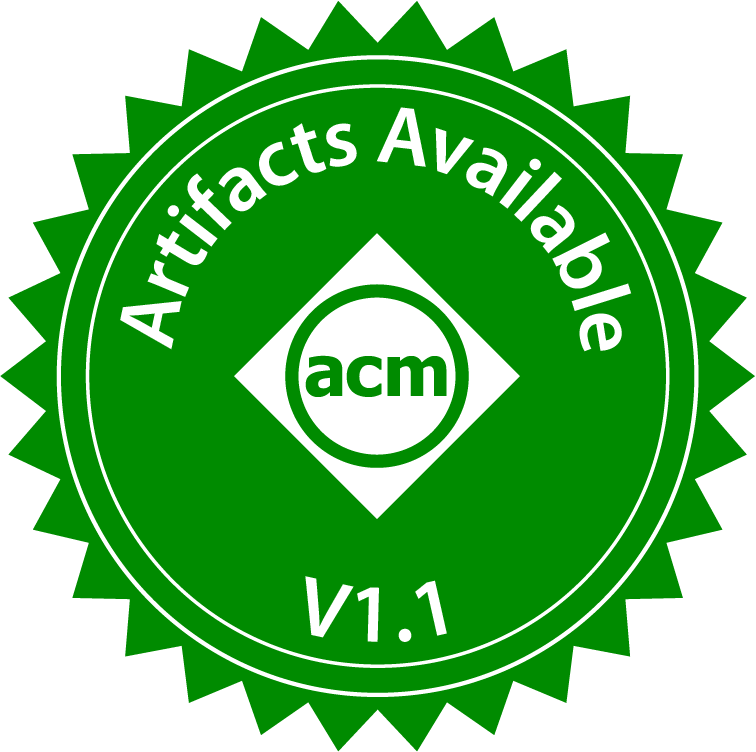}
\acmBadgeR[https://sysartifacts.github.io/atc2026/results]{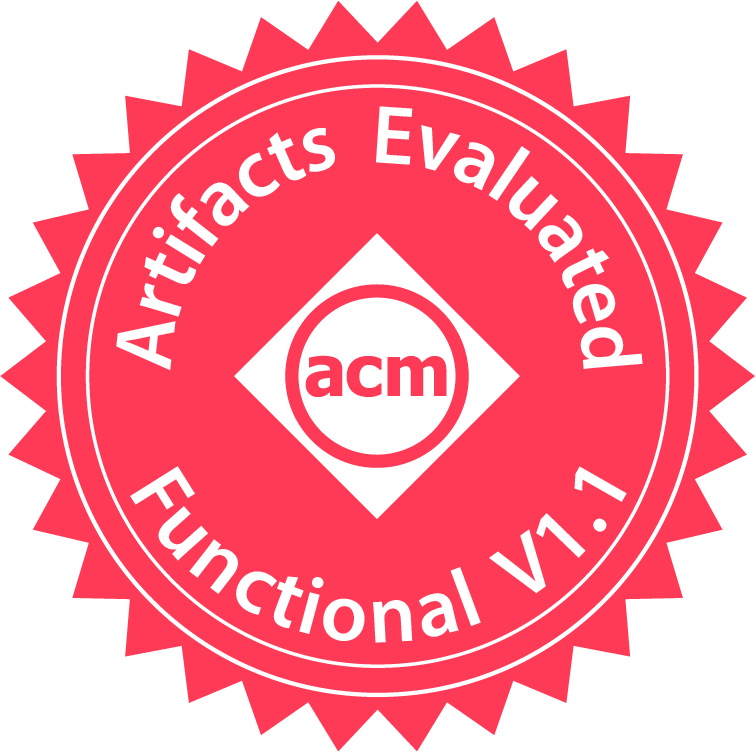}
\acmBadgeR[https://sysartifacts.github.io/atc2026/results]{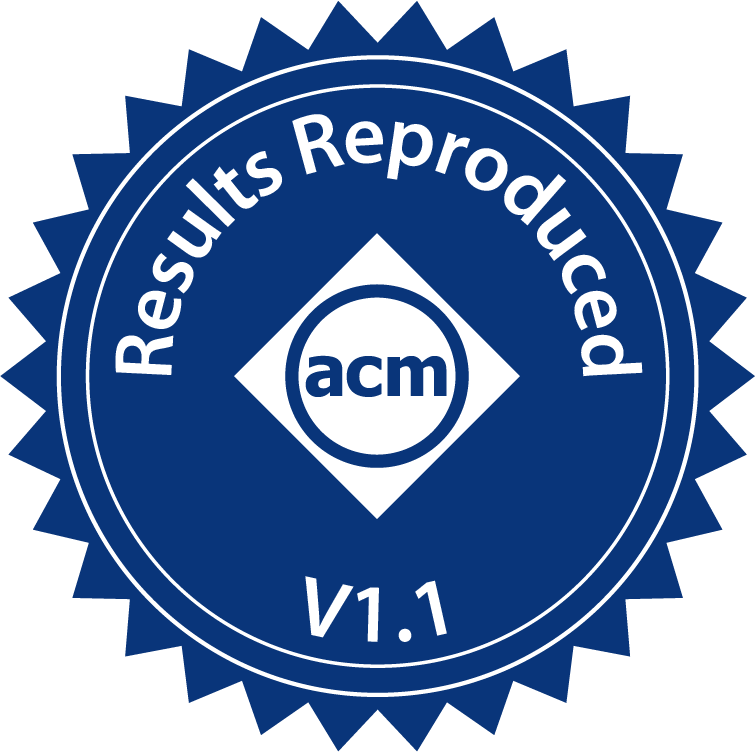}

\begin{document}
\pagestyle{empty}

\title{\sys: Expert-Locality-Aware Decode Routing for PD-Disaggregated MoE Serving}

\author{Sangjin Choi}
\authornote{Work partially done during an internship at Microsoft Research.}
\affiliation{%
  \institution{KAIST}
  \city{Daejeon}
  \country{Republic of Korea}
}
\email{sjchoi@casys.kaist.ac.kr}

\author{Sukmin Cho}
\authornotemark[1]
\affiliation{%
  \institution{KAIST}
  \city{Daejeon}
  \country{Republic of Korea}
}
\email{smcho@casys.kaist.ac.kr}

\author{Yifan Xiong}
\affiliation{%
  \institution{Microsoft Research}
  \city{Beijing}
  \country{China}
}
\email{yifan.xiong@microsoft.com}

\author{Ziyue Yang}
\authornote{Work partially done while the author was at Microsoft Research.}
\affiliation{%
  \institution{Shanghai Xingyunzhili Artificial Intelligence Institute}
  \city{Shanghai}
  \country{China}
}
\email{ziyue.yang@xingyunzhili.com}

\author{Youngjin Kwon}
\authornote{Corresponding author.}
\affiliation{%
  \institution{KAIST}
  \city{Daejeon}
  \country{Republic of Korea}
}
\email{yjkwon@kaist.ac.kr}

\author{Peng Cheng}
\affiliation{%
  \institution{Microsoft Research}
  \city{Redmond}
  \state{WA}
  \country{USA}
}
\email{pengc@microsoft.com}

\begin{abstract}
Equal load does not mean equal decode latency in prefill--decode (PD)
disaggregated mixture-of-experts (MoE) serving.
In bandwidth-bound decode, more distinct experts mean more weight loads and higher latency.
We present \sys, a lossless, expert-locality-aware decode router.
Prefill signatures predict decode expert usage.
Balanced $K$-means fits per-decoder centroids with background refitting.
Locality-band routing selects the least-loaded worker among close matches.
A block-granular cache
keeps signatures exact under prefix caching.
We evaluate \sys in vLLM on up to 40 GPUs.
Across three MoE models on 12 public task benchmarks and multilingual
WildChat, it reduces median time-per-output-token (TPOT) by 5.4--14.5\% on average
across request rates and up to 17.8\% over the best of four load-balancing baselines,
preserving model outputs and token-to-expert assignments.
\end{abstract}



\maketitle

\section{Introduction}
\label{sec:introduction}

Large language model (LLM) serving is moving toward prefill--decode (PD)
disaggregation, which runs prompt processing (prefill) and token generation
(decode) on separate worker pools~\cite{distserve}. Prefill processes prompt
tokens in parallel and is compute-bound,
while decode generates tokens sequentially and is memory-bandwidth-bound.
Colocating them causes phase interference---long prefills stall ongoing
decoding---and couples their resource allocation. An $x$P$y$D
deployment instead provisions $x$ prefill and $y$ decode workers
separately~\cite{dynamo,llmd,deepseekv3,mooncake}, turning routing into the
central serving decision: each request is assigned a prefill worker for its
prompt and, after prefill, a decode worker for its
generation. Prefill-side routing has drawn most of the attention:
cache-aware policies route requests to improve prefix-cache reuse~\cite{preble}. Decode-side routing has
not: existing policies balance load and otherwise treat decode workers as
interchangeable. For dense models they are, since equal-load workers do equal
feed-forward work. For mixture-of-experts (MoE) models they are not.

For MoE models, equal load does not mean equal latency. Decode is
memory-bandwidth-bound, and during batched decoding its cost is set by the \emph{union}
of distinct experts the batch loads from HBM each step. Here sparsity inverts.
The property that makes MoE cheap, routing each token to only a few
experts~\cite{qwen3,gptoss,gemma4}, fragments the decode batch across
experts and destroys the weight reuse a dense batch enjoys, where one weight load
amortizes over every token. A step pays for every expert any of its tokens
selects. Within a decode batch, overlapping expert choices enable weight reuse,
whereas different choices enlarge the union. Workers
with equally many requests can incur different expert-weight
traffic, even when every token selects the same number of experts.
On Qwen3-30B-A3B, growing the active-expert count from 16 to 128 raises MoE-layer
latency $4.7\times$ at fixed batch size, while batch size at fixed active-expert
count barely moves it (\cref{sec:motivation_active_experts}).
Because request placement determines which requests can be batched together,
expert composition is a first-order latency knob that load-based routing
cannot see. This motivates a second routing axis: \emph{expert locality}.

Expert locality is exploitable because it is structured by domain.
The MoE gate%
\footnote{We call the MoE router the MoE gate to avoid confusion with the PD router.}
picks each
token's experts from its hidden representation, so requests from the same domain
activate overlapping experts: across tasks, code, math, medical, and legal
prompts exercise distinct expert regions, and multilingual traffic separates by
language. Expert choice is thus correlated across related requests.
Concretely, we observe same-domain decode batches activate
$17$--$21\%$ fewer distinct experts per step than mixed batches on task workloads
(\cref{sec:motivation_domains}), so routing similar requests to the same decode
worker can shrink its per-step expert union, while load-only routing scatters them.

Structure helps only if the router can see it in time. Placement happens at the
prefill-to-decode handoff, before decoding begins, so the request's decode-time expert choices
are not yet observable. But prefill has already pushed the prompt through the same
gates that will route decode. Measurements on a mixture of 12 public task
benchmarks and real-world multilingual WildChat~\cite{wildchat} traffic show substantial
agreement: across the three evaluated MoE models, the mean per-request prefill--decode
correlation is $0.69$, compared with $0.18$ for randomly mismatched requests
(\cref{sec:motivation_prefill_decode}). Expert locality is therefore not only
structured but visible when the router must act.

We present \sys, a lossless, expert-locality-aware decode router for
PD-disaggregated MoE serving. \sys uses \emph{expert signatures} derived from
each request's prefill expert activations to route requests across decode workers, grouping
requests predicted to share experts during decoding. These signatures capture
variation in expert usage even within a single domain, without requiring domain
labels for individual requests. All decode workers run the same model, and \sys changes only request
placement across decode workers,
preserving model weights and gate-selected expert execution. Inaccurate
predictions of decode-time expert overlap may reduce locality benefits, but model
outputs remain unchanged. To our knowledge, \sys is the first to reduce the
active-expert union through lossless decode routing in PD-disaggregated MoE
serving.

Routing must satisfy two objectives that need different information.
Locality reflects expert-usage patterns across requests, while load reflects
the requests currently assigned to each worker. Pure
locality overloads workers serving popular domains; pure load balancing scatters expert-similar
requests. \sys combines balanced clustering and locality-band routing.
\emph{Balanced $K$-means} initializes one locality region per decode worker,
capturing structure while balancing cluster sizes. During serving,
centroids are periodically refitted from recent signatures in the background
while requests continue using the current centroids.
Even balanced clusters cannot guarantee balanced runtime loads when arrivals
favor one cluster or requests finish at different times.
\mbox{\emph{Locality-band routing}} identifies candidate workers whose centroids fall
within a similarity band of the request's best match and selects the least-loaded among
them at runtime. The band preserves locality, while the load-based choice responds
to transient imbalance.

One case stresses the online path: prefix caching. A cache hit skips
prefill for the shared prefix, so the gate never runs there and the
signature is incomplete---worst exactly for the cache-hit requests
that prefix-aware prefill routing concentrates. \sys therefore keeps an
expert-signature cache co-indexed with the KV cache at block
granularity: each block carries its tokens' expert footprint, and
summing the counts from cached and freshly computed blocks recovers the full
signature. The signature cache follows the lifecycle of KV-cache blocks, remaining
coherent across partial hits, evictions, and block reuse without recomputing
cached tokens or changing cache management.

We implement \sys in vLLM~\cite{vllm}: prefill-time signature capture, centroid fitting,
locality-band routing, and the block-granular signature cache.\footnote{Source code:
\url{https://doi.org/10.5281/zenodo.23310788}.} We evaluate three MoE models
(Qwen3-30B-A3B~\cite{qwen3}, GPT-OSS-120B~\cite{gptoss},
Gemma-4-26B-A4B~\cite{gemma4}) on two complementary workload families:
a mixture of 12 public benchmarks spanning code, math, medicine, and law
(\emph{Task}), and $14{,}000$ real-world multilingual prompts from
WildChat (\emph{Language}).
Round-robin (RR) has the lowest median time-per-output-token (TPOT),
averaged across request rates, among our load-balancing baselines.
Relative to RR, \sys reduces median TPOT by $8.6$--$14.5\%$ on Task
and $5.4$--$9.8\%$ on Language on average across request rates,
with peak reductions of $17.8\%$ and $11.9\%$, respectively
(\cref{sec:eval_main}). Median time-to-first-token (TTFT) remains largely unchanged.
The signature cache is under $1\%$ of the
KV cache (\cref{sec:design_prefix_cache}).
\sys generalizes to a 235B deployment spanning 40 GPUs with expert parallelism (EP)
(\cref{sec:eval_generalization}).

\sys makes three contributions:

\begin{itemize}[leftmargin=*, itemsep=0pt, topsep=0pt]
\item \textbf{Expert locality as a predictable routing axis.} Even at equal
worker load, MoE decode latency varies with the number of distinct experts
activated in each decode batch. This locality is
structured by domain and, crucially, observable from prefill before decode begins.

\item \textbf{Expert-locality-aware decode routing.} \sys builds
prefill-derived expert signatures that predict decode-time expert overlap,
partitions them across decoders with
balanced $K$-means, and routes with a locality band that balances expert
locality against live load.

\item \textbf{Prefix-cache-coherent signatures.} A block-granular signature cache
co-indexed with the KV cache keeps signatures exact across partial hits, full
hits, and evictions, so \sys composes with existing prefix caching mechanisms
at negligible cost.
\end{itemize}

\section{Background}
\label{sec:background}

\subsection{Mixture-of-Experts Large Language Models}
\label{sec:background_moe}

\looseness=-1
Mixture-of-experts has become a primary architecture for scaling frontier large language models, decoupling model capacity from per-token computation~\cite{qwen3}. Instead of applying the same dense feed-forward network (FFN) to every token, an MoE layer replaces the FFN with multiple expert FFNs and an MoE gate that activates a small subset of experts for each token. The model can thus contain many more parameters than a dense model while using only a fraction per forward pass. This sparse activation makes MoE
attractive for efficient scaling: it increases model capacity and enables richer
expert specialization without proportionally increasing per-token FLOPs. Recent
architectures push this idea further with fine-grained experts~\cite{deepseekv3}, splitting
FFN capacity across more, smaller experts to route each token to a
more specialized expert combination while keeping per-token FLOPs bounded.

\subsection{Prefill--Decode Disaggregated Serving}
\label{sec:background_pd}

Prefill--decode disaggregation~\cite{distserve} is a practical
architecture for latency-sensitive LLM serving. Inference has two
phases: \emph{prefill} processes the prompt in parallel and is
compute-bound, while \emph{decode} generates tokens autoregressively and
is latency-sensitive and memory-bandwidth-bound. Colocating them causes
phase interference---long prefills stall latency-sensitive decode
steps---and forces one resource allocation onto two mismatched
workloads. PD disaggregation separates the phases onto independent
pools, typically an \emph{xPyD} configuration of $x$ prefill and $y$
decode workers, scaling capacity per phase.

This makes routing a central serving decision. A \emph{PD router}
assigns each request prefill and decode workers, determining the
destination of its KV-cache transfer.
Cache-aware routers select prefill workers to
improve prefix-cache reuse~\cite{mooncake}. Load-balancing policies such
as RR, join-the-shortest-queue (JSQ), and power-of-two-choices
(P2C) spread requests across workers. These policies do not account for expert
overlap among requests assigned to the same decode worker.

\subsection{Prefix Caching}
\label{sec:background_prefix}

The KV cache stores attention keys and values for processed tokens.
Prefix caching reuses these values for matching prompt prefixes,
avoiding repeated prefill computation~\cite{sglang}.

PagedAttention~\cite{vllm} applies the idea of virtual-memory paging to the
KV cache. It divides each request's KV cache into fixed-size
token blocks that can be stored non-contiguously in GPU memory. Prefix
caching reuses matching leading blocks across requests, so a \emph{partial hit}
requires computing only the remaining prompt tokens. Cached blocks may
outlive the requests that created them; once no request uses a block, it
may be evicted and its storage reused.

\section{Motivation}
\label{sec:motivation}

\subsection{Active Expert Count Drives MoE Decode Latency}
\label{sec:motivation_active_experts}

MoE decode latency is governed by the number of distinct experts
activated at each decode step. Sparsity reduces computation but
amplifies decode's memory-bandwidth bottleneck by fragmenting the
batch across experts. In a dense model, every token in a decode batch
executes the same FFN, so one weight load amortizes over the whole
batch. In an MoE model, the gate partitions the batch among experts,
and weight reuse occurs only among tokens sharing an expert; as
experts grow more numerous and finer-grained, fragmentation lowers
arithmetic intensity further. MoE decode cost is therefore dominated
by expert weight access: each step fetches the weights of every expert
the batch activates.

\begin{figure}[!t]
\centering
\includegraphics[width=0.95\columnwidth]{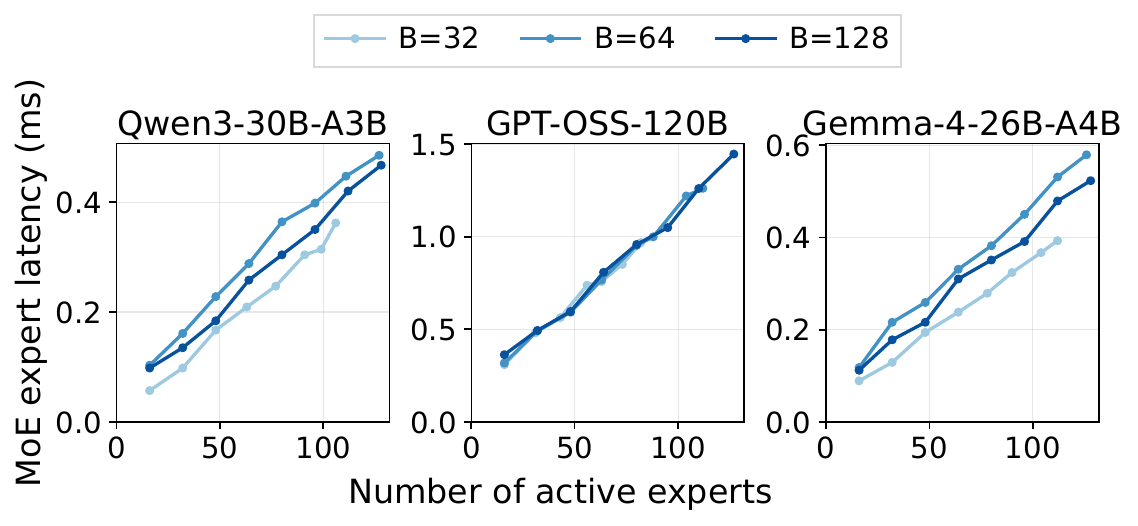}
\vspace{-1em}
\caption{MoE layer latency scales with active experts, not batch
size (single MoE layer, one MI300X).}
\Description{Three line charts for Qwen3-30B-A3B, GPT-OSS-120B, and
Gemma-4-26B-A4B plot MoE layer latency against the number of active
experts, with separate curves for batch sizes 32, 64, and 128.
Latency rises with active-expert count, while the curves for different
batch sizes remain close.}
\label{fig:moe_latency}
\end{figure}

We benchmark single-layer MoE expert computation on one MI300X
(\Cref{fig:moe_latency}). Across the three models, varying batch size
at fixed active-expert count has little effect. By contrast,
increasing active experts from 16 to 128 raises Qwen3-30B-A3B latency
$4.7\times$ at batch size 64.

This result shows that MoE decode efficiency depends on expert reuse
within a batch. However, exploiting this requires the serving system to
know which requests are likely to activate overlapping experts. We next
show that this overlap is not random. Instead, expert usage is structured
by request domain, making expert locality a predictable property that a
serving system can exploit.

\begin{figure*}[!t]
\centering
\includegraphics[width=\textwidth]{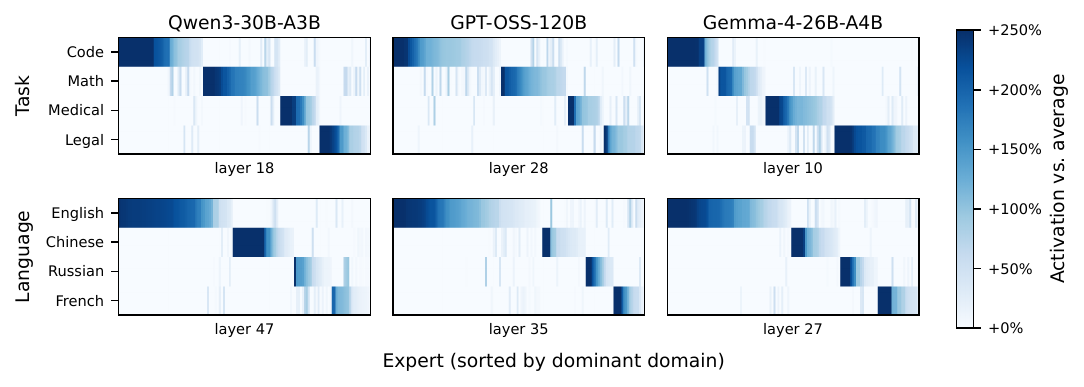}
\vspace{-2.2em}
\caption{Decode-phase per-expert activation relative to the cross-domain mean,
for three MoE models along task (top) and language (bottom,
WildChat~\cite{wildchat}) domains at each model's most discriminative layer.
Darker is above-average (below-average clipped to white); experts are reordered
per panel into contiguous per-domain blocks. Each domain over-activates a
distinct subset of experts. Task domains:
Code~\cite{humaneval, bigcodebench, ds1000, mbpp},
Math~\cite{gsm8k,math500,olympiadbench,aquarat},
Medical~\cite{medqa, pubmedqa, mmlu}, and
Legal~\cite{lexglue}.}
\Description{Six heatmaps compare expert activation across three models.
The upper row shows Code, Math, Medical, and Legal domains; the lower
row shows English, Chinese, Russian, and French. Within each heatmap,
columns are experts grouped by their dominant domain. Darker cells
indicate higher activation relative to the cross-domain mean, while
below-average values are white. Each domain has a distinct region of
above-average expert activation.}
\label{fig:domain}
\end{figure*}

\begin{samepage}
\subsection{Expert Locality Is Structured by Domain}
\label{sec:motivation_domains}

Expert locality is exploitable because it is structured: requests from
the same domain tend to activate overlapping expert subsets. The MoE
gate selects experts using each token's hidden representation, which
encodes contextual features of the request. We study task and language
as two examples of the request characteristics that shape expert usage.

\end{samepage}

\begin{figure}[t]
\centering
\includegraphics[width=0.95\columnwidth]{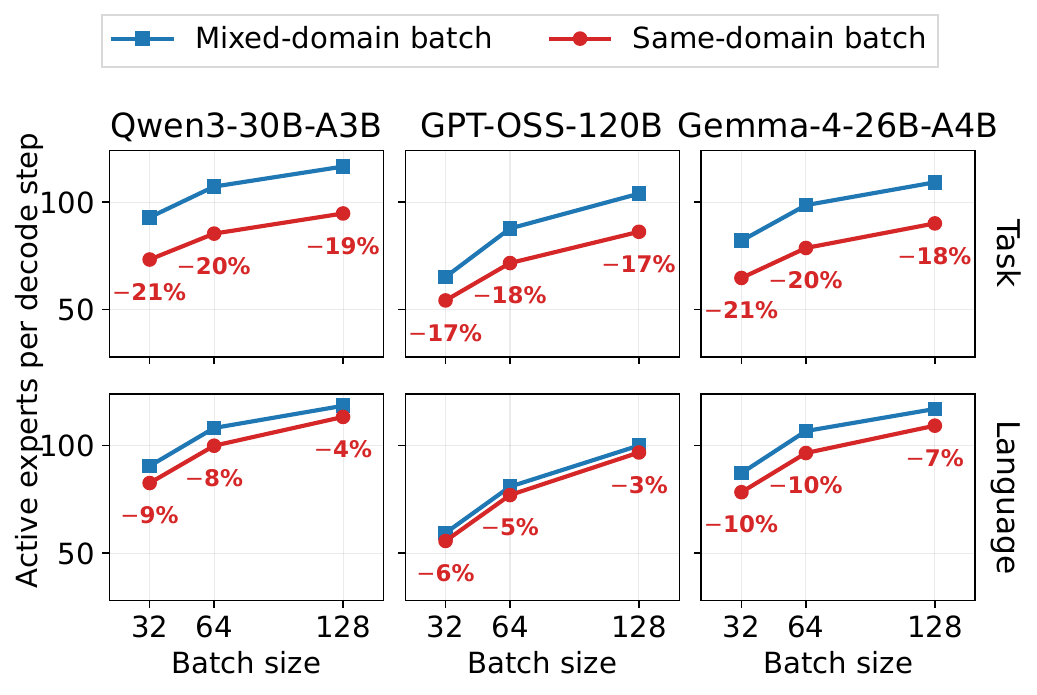}
\vspace{-1em}
\caption{Same-domain batches (red) activate fewer experts per decode
step than mixed-domain (blue),
for task (top) and language (bottom) across three MoE models.}
\Description{Six line charts show active experts per decode step at
batch sizes 32, 64, and 128. Columns correspond to Qwen, GPT-OSS, and
Gemma; rows correspond to Task and Language. The red same-domain
curves lie below the blue mixed-domain curves at every tested batch
size. Reductions range from 17 to 21 percent for Task and 3 to
10 percent for Language. Both curves rise as batch size increases.}
\label{fig:opp_sweep}
\end{figure}

\Cref{fig:domain} shows decode-phase expert activation across three
models: Qwen3-30B-A3B, GPT-OSS-120B, and Gemma-4-26B-A4B. The task axis uses a
mixture of 12 public benchmarks spanning code, math, medicine, and law.
The language axis uses multilingual requests from WildChat, a corpus of one million real-world
user--ChatGPT interactions spanning 68 languages such as English, Chinese,
Russian, and French. In each heatmap, rows correspond to domains and
columns correspond to experts, with color indicating activation
relative to the average for that expert across domains. Experts are sorted by
their dominant domain, so experts most activated by the same domain appear
contiguously.

\begin{figure}[t!]
\centering
\includegraphics[width=0.95\columnwidth]{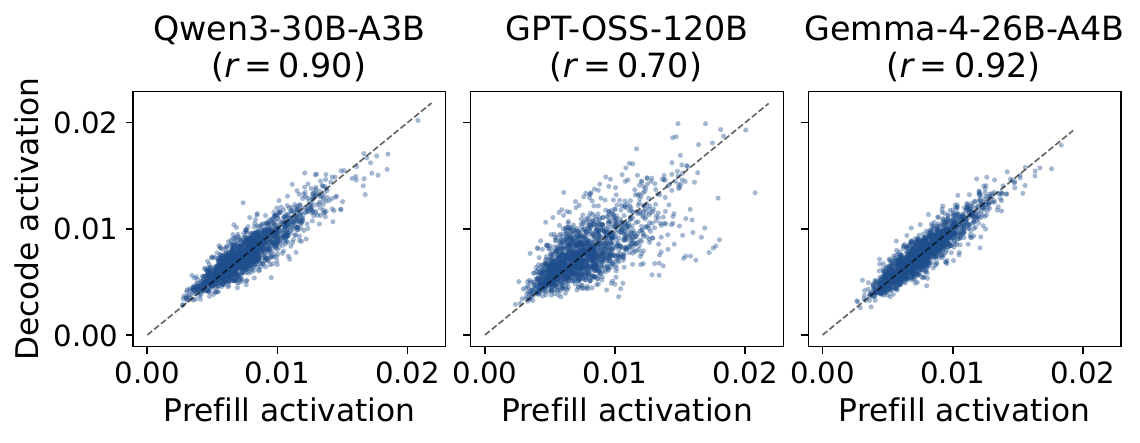}
\vspace{-1em}
\caption{Prefill--decode correlation in domain-pooled activation profiles.
Each point represents an expert index within a domain; activations
are summed over requests and layers, then normalized within each
domain and phase.}
\Description{Three scatter plots compare normalized prefill expert activation
on the horizontal axis with normalized decode expert activation on the
vertical axis. Each point is an expert within a pooled domain.
The points follow a positive diagonal trend. Correlations are 0.90
for Qwen3-30B-A3B, 0.70 for GPT-OSS-120B, and 0.92 for Gemma-4-26B-A4B.}
\label{fig:pd_scatter}
\end{figure}

Across models, expert activation is strongly structured by both task and
language. Code, math, medical, and legal requests over-activate
different expert subsets. The same pattern appears on the multilingual axis:
English, Chinese, Russian, and French requests activate different expert
regions rather than uniformly exercising the full expert pool. This shows that
expert selection is correlated across related requests.

This structure creates an opportunity to exploit expert locality through
request placement. \Cref{fig:opp_sweep} quantifies the benefit: at matched
batch sizes of 32--128, same-domain batches activate 17--21\% (Task) and
3--10\% (Language) fewer experts per step than mixed-domain batches.
The reduction comes from grouping requests with overlapping expert usage,
not from serving fewer requests.

\subsection{Prefill Predicts Decode Expert Activation}
\label{sec:motivation_prefill_decode}

Structure helps only if the router can see it in time. At the
prefill--decode handoff, the router must assign each request to a decode
worker before observing its generation-time expert usage. Prefill,
however, has already processed the prompt through the same MoE gates,
exposing a potential signal of which experts the request will use during
decoding.

\Cref{fig:pd_scatter} compares prefill and decode expert activation across
the task and language domains studied above. For each domain, we aggregate
expert activations across requests and layers to obtain a normalized
activation profile for each phase. The $x$-axis shows normalized prefill
activation, and the $y$-axis shows
normalized decode activation; points near the diagonal indicate comparable
usage in both phases. The correlation is strong for Qwen3-30B-A3B and
Gemma-4-26B-A4B, and substantial for GPT-OSS-120B.
This agreement also holds for individual requests. We compute Pearson
correlations over layer--expert pairs using
$1{,}000$ requests per model--workload setting. Across the six
model--workload settings, the mean per-request Pearson correlation ranges
from $0.56$ to $0.79$, averaging $0.69$. Randomly mismatched prefill--decode
pairs within the same setting yield a mean correlation of $0.18$.

Expert locality is therefore not only structured but visible when the
router must act. Prefill expert activations let the router identify requests
likely to share experts during decoding and place them on the same decode
worker. To guide this placement, \sys summarizes each request's prefill
expert activations in an \emph{expert signature}.

\subsection{Challenges}
\label{sec:motivation_challenges}

Turning this opportunity into a practical routing policy raises three
challenges.

\aptLtoX{}{\vspace{-0.25\baselineskip plus -1pt minus -0.1pt}}
\subsubsection{Designing the Expert Signature}
The challenge is to map prefill expert activations to a signature space
in which proximity reflects how much requests overlap in decode-time
expert activation. This representation is the building block of \sys's
design: every routing decision derives from it, and the locality benefit
depends on how well it preserves this relationship.
The design space is large: raw
activation counts, gate logits, layer masks, and reweighting
schemes all yield candidate representations, each inducing a
different geometry. Selecting among them requires a criterion
that measures how faithfully signature proximity predicts
decode-time expert overlap, independent of any particular
clustering or routing policy built on top.
Otherwise, improvements from a better representation could be confused
with effects of the clustering method or the online routing rule.

\aptLtoX{}{\vspace{-0.25\baselineskip plus -1pt minus -0.1pt}}
\subsubsection{Reconciling Locality and Load}
\label{sec:challenge_load}
\begin{figure}[t]
\centering
\includegraphics[width=0.95\columnwidth]{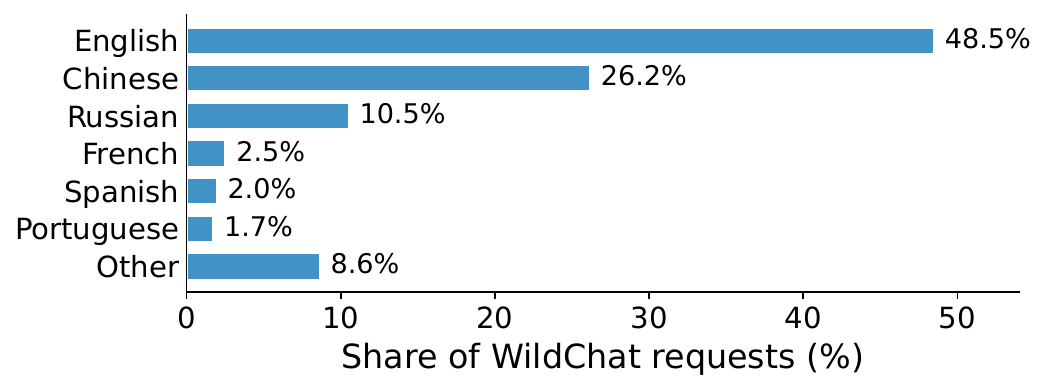}
\vspace{-1em}
\caption{Language workload (14,000 WildChat~\cite{wildchat} prompts): English
and Chinese account for $\sim$75\% of requests.
}
\Description{A horizontal bar chart shows language shares in the
14,000-prompt WildChat workload, including 1,000 calibration and
13,000 evaluation prompts. English accounts for 48.5 percent, Chinese
26.2, Russian 10.5, French 2.5, Spanish 2.0, and Portuguese 1.7.
All other languages together account for 8.6 percent. English and
Chinese dominate the request volume.}
\label{fig:wildchat}
\end{figure}

Locality alone is not enough. Non-MoE compute---attention, QKV
projection, and normalization---scales with active requests, not their
shared experts. Locality-only routing concentrates traffic on workers
matching popular domains; English and Chinese account for
$\sim$75\% of our Language workload (\Cref{fig:wildchat}). The resulting overload
increases non-MoE work and tail TPOT. Conversely, round-robin or pure
JSQ spreads requests but scatters expert-similar ones, erasing locality
benefits. Routing must preserve locality while balancing load.

Two sources of imbalance make this trade-off difficult. First, some
expert-usage patterns are more common than others, so grouping similar
requests can assign too much traffic to a few workers. Second, even a
grouping balanced over past requests does not guarantee balanced queues
at every moment: bursts of arrivals can favor one cluster, while
completions drain different workers at different rates. Routing must account for
both the workload distribution and current worker loads while
preserving expert locality.

\aptLtoX{}{\vspace{-0.25\baselineskip plus -1pt minus -0.1pt}}
\subsubsection{Prefix-Cache Signature Coherence}
\label{sec:challenge_prefix}

Cache hits (\cref{sec:background_prefix}) skip MoE gate execution for
cached tokens. Their expert activations were produced and discarded during
an earlier request's prefill, leaving this request's observed expert
footprint incomplete and potentially leading to poorer routing decisions.
The same prompt can then produce different signatures depending on how
much of its prefix is cached. The signature should represent the full
prompt's expert usage, regardless of which tokens require fresh prefill.

Caching a whole-prompt signature does not solve this problem.
It combines activations from the entire
prompt, whereas the new request needs only the shared prefix's activations
to combine with those of newly computed tokens.
Cached KV blocks can also be evicted and their storage
reused for new tokens. The challenge is to recover the full prompt's
expert footprint across partial hits while maintaining signature
coherence throughout the lifecycle of KV-cache blocks, without disabling
the cache or re-running prefills.

\section{Design}
\label{sec:design}

We present \sys, an \textbf{E}xpert-\textbf{L}ocality-aware
\textbf{D}ecode \textbf{R}outing policy for PD-disaggregated MoE
serving. Instead of placing decode requests by load alone, \sys
uses each request's prefill expert activations to choose a decoder
where it is likely to share experts with other requests, reducing TPOT.

\subsection{Overview}
\label{sec:design_overview}

\begin{figure}[t]
\centering
\includegraphics[width=0.95\columnwidth]{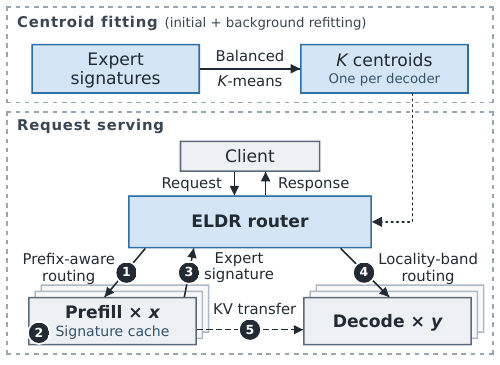}
\vspace{-1.0em}
\caption{\sys architecture. Prefill workers collect expert activations
in a block-granular signature cache, and the router uses the resulting
request signatures for locality-band decode routing. Balanced $K$-means
initializes one centroid per decoder from calibration signatures and
refreshes them from recent signatures in the background.}
\Description{An architecture diagram separates centroid fitting from
request serving. Above, balanced K-means converts expert signatures
into one centroid per decoder, for initial fitting and background
refitting. A dotted arrow supplies centroids to the ELDR router.
Below, the router sends client requests to prefill workers using
prefix-aware routing. Prefill workers maintain a signature cache and
return expert counts to the router, which uses locality-band routing
to select a decoder. KV caches transfer from prefill to decode
workers, and responses return to the client through the router.}
\label{fig:eldr_overview}
\end{figure}

\Cref{fig:eldr_overview} shows the \sys architecture.
\sys adds a router and a prefill hook to a PD-disaggregated
serving stack. The hook records expert-activation counts at KV-block
granularity in a signature cache. At the prefill--decode handoff, the router transforms
these counts into a request signature and selects a decoder using the
current centroids. We implement \sys on vLLM~\cite{vllm} as a lossless
routing layer that changes request placement across decode workers while
preserving model weights and gate-selected expert execution.

\aptLtoX{}{\vspace{-0.5\baselineskip plus -2pt minus -0.2pt}}
\paragraph{Expert Signature}\aptLtoX{}{\hspace{0.5pt}}%
\sys builds an \emph{expert signature} (\cref{sec:design_signature})
from each request's prefill expert activations to predict its decode-time
expert usage (\cref{sec:motivation_prefill_decode}). Cosine similarity
between signatures estimates how much two requests' decode-time expert
usage overlaps. Routing requests with similar signatures to the same decode
worker can reduce the number of distinct experts activated within decode batches.

\aptLtoX{}{\vspace{-0.5\baselineskip plus -2pt minus -0.2pt}}
\paragraph{Centroid Fitting}\aptLtoX{}{\hspace{0.5pt}}%
Before serving, \sys fits the signature transform using calibration
requests. Balanced
clustering (\cref{sec:design_clustering}) groups similar signatures for
locality and balances cluster sizes, with one centroid per worker. During serving,
\sys periodically refits centroids from recent signatures to adapt to
workload drift. It matches new centroids to current ones so each decoder
retains a similar signature region.
Refitting and matching run in the background on the CPU while requests
use the current centroids. Once matching completes, the router
atomically switches to the updated centroids.

\aptLtoX{}{\vspace{-0.5\baselineskip plus -2pt minus -0.2pt}}
\paragraph{Request Serving}\aptLtoX{}{\hspace{0.5pt}}%
During serving, \sys leaves the standard PD pipeline intact and
adds two lightweight steps: expert-count collection during prefill and
a routing decision at the prefill--decode handoff:

\begin{enumerate}[leftmargin=*, topsep=0pt]
\item The router sends the request to a prefill worker using
prefix-aware routing.

\item During prefill, \sys records expert-activation counts at
KV-block granularity in a signature cache co-indexed with the KV cache
(\cref{sec:design_prefix_cache}).

\item At handoff, the prefill worker sums the request's per-block
counts, including those reused from cached prefix blocks, and returns
the resulting counts to the router with the KV-transfer metadata.

\item The router transforms these counts into the request's expert
signature (\cref{sec:design_signature}), then applies
\emph{locality-band routing} (\cref{sec:design_route}): it compares
the signature against the decoder centroids and routes to the
least-loaded worker within the locality band of the closest centroid.

\item The selected decoder pulls the KV cache and generates on the
unchanged decode engine. Routing similar-signature requests to the same
worker can shrink the active-expert union within decode batches.
\end{enumerate}

\subsection{Expert Signature}
\label{sec:design_signature}

\sys builds each request's expert signature from
expert-activation counts collected during prefill. Raw counts
alone can obscure differences between requests: common experts dominate
the counts, while some layers provide little information about decode-time
expert usage. \sys therefore reweights the counts and selects informative
layers before normalizing the signature.

\aptLtoX{}{\vspace{-0.25\baselineskip plus -1pt minus -0.1pt}}
\subsubsection{Goal of Signature Design}
\label{sec:sig_metric}

\sys uses signature similarity to group requests likely to
share experts during decoding. Prefill signature similarity should
therefore reflect similarity in
decode-time expert usage.

Let $s_i$ denote the prefill signature of request $i$. To describe its
decode-time pattern, we record, for each layer $\ell$ and expert $e$,
the fraction of decode steps in which that expert is selected. We
flatten these fractions across all $(\ell, e)$ pairs and L2-normalize
the resulting vector to obtain $p_i$.

For randomly sampled calibration request pairs $(i,j)$, we compute the
cosine distance between $s_i$ and $s_j$, and separately between $p_i$
and $p_j$. Signature quality $\rho$ is the Spearman rank
correlation between these two distance sets:
\begin{equation}
\rho \;=\; \mathrm{Spearman}\big(\,
  \mathrm{cos\text{-}dist}(s_i, s_j),\;\mathrm{cos\text{-}dist}(p_i, p_j)
\,\big).
\label{eq:rho}
\end{equation}

Spearman correlation compares distance ordering without
assuming a linear relationship between their values. A high $\rho$
indicates that request pairs judged more similar by their prefill
signatures also tend to have more similar decode-time expert-activation
patterns.
This evaluates signature quality independently of clustering and
routing policies.

\aptLtoX{}{\vspace{-0.25\baselineskip plus -1pt minus -0.1pt}}
\subsubsection{Signature Design}
\label{sec:sig_construction}
Four steps turn prefill expert counts into a signature $s_r$.
Calibration fixes the IDF weights and layer mask. At handoff, the
router applies both to the counts and L2-normalizes the result.

\begin{figure}[t]
\centering
\includegraphics[width=0.95\columnwidth]{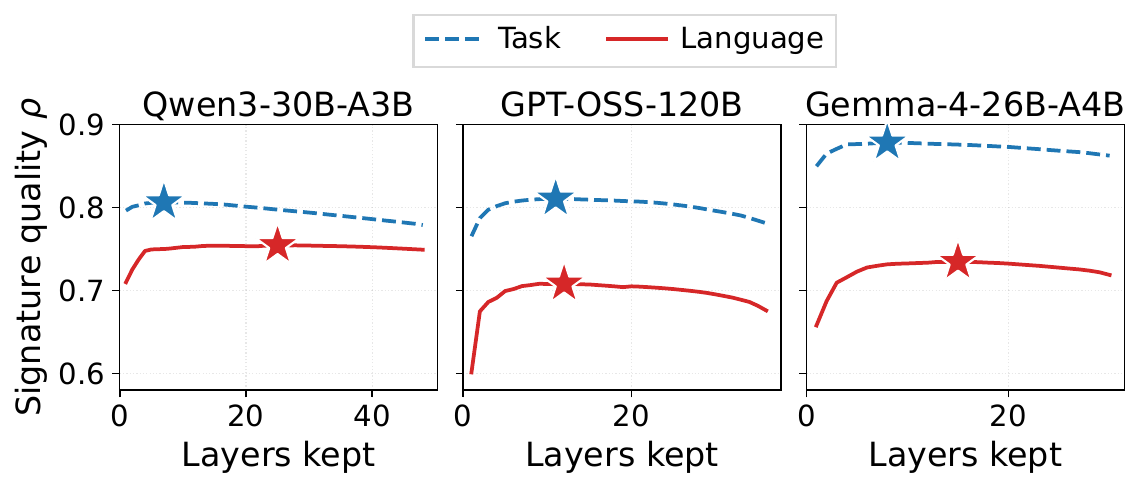}
\vspace{-1em}
\caption{Signature quality~$\rho$ (Eq.~\ref{eq:rho}) versus the number of
layers kept under greedy layer selection. One panel per model;
task (blue dashed) and language (red solid). The star
marks the peak~$N^{*}$ chosen by \sys{}'s offline fit.}
\Description{Three line charts plot signature quality rho against
the number of layers retained for Qwen, GPT-OSS, and Gemma. Task
curves are blue dashed lines and Language curves are red solid lines. Stars mark the selected
peak of each curve. In all six settings, the peak occurs before all
layers are included; adding further layers lowers signature quality.}
\label{fig:sig_layers}
\end{figure}

\begin{enumerate}[leftmargin=*]
\item \emph{Count.} For request $r$, let $c_r(\ell,e)$ be the number
of prefill tokens routed to expert $e$ at layer $\ell$. These counts
record discrete top-$k$ selections, matching the expert activations
measured during decode. Gate probabilities, in contrast, also assign
weight to experts outside the selected top-$k$.

\item \emph{Downweight common experts.} Experts that fire on nearly
every request provide little information for distinguishing requests.
We weight each count by its inverse document frequency (IDF),
$w(\ell,e)=\log\frac{|\mathcal{C}|+1}{\mathrm{df}(\ell,e)+1}$,
where $\mathcal{C}$ is the calibration set and $\mathrm{df}(\ell,e)$
counts requests in which expert $e$ fires at least once at layer $\ell$.
The weighted count, $\tilde c_r(\ell,e)=c_r(\ell,e)\,w(\ell,e)$,
gives rare experts more influence relative to common ones.

\item \emph{Select layers.} Some layers add dimensions without
helping predict decode-time expert usage. During calibration, we rank
the model's $L$ layers by greedy selection: starting from an empty set,
we repeatedly add the remaining layer that gives the highest $\rho$
when combined with those already selected. We keep the first $N^{*}$
layers, where $\rho$ peaks (the~$\bigstar$ in~\Cref{fig:sig_layers}),
forming the layer mask $\mathcal{S}$. At runtime, we concatenate the
weighted counts from these layers into a single vector $x_r$.

\item \emph{Normalize.} We L2-normalize $x_r$ to obtain
$s_r=x_r/\lVert x_r\rVert_2$, so similarity reflects the shape of expert
usage rather than the request's token count.
\end{enumerate}

\begin{table}[t!]
\centering
\caption{Mean signature quality~$\rho$ (Eq.~\ref{eq:rho}) for four
signature transformations, equally averaged over three models and two workloads.}
\label{tab:sig_transform}

\setlength{\tabcolsep}{2.25pt}
\begin{tabular}{@{}l@{\hspace{1em}}cccc@{}}
\toprule
Transform & Binary & Gate-prob & Count & Count$\cdot$idf \\
\midrule
Mean $\rho$ & 0.470 & 0.673 & 0.749 & \textbf{0.764} \\
\bottomrule
\end{tabular}

\end{table}

\aptLtoX{}{\vspace{-0.25\baselineskip plus -1pt minus -0.1pt}}
\subsubsection{Validation}
\label{sec:sig_validation}
We evaluate signature transforms and layer selection using $\rho$
(Eq.~\ref{eq:rho}) on 1{,}000 calibration requests for each of the six
model--workload settings.
\cref{sec:design-validation} compares count$\cdot$idf and gate-probability
signatures in end-to-end serving with greedy layer selection,
balanced clustering, and locality-band routing.

\aptLtoX{}{\vspace{-0.5\baselineskip plus -2pt minus -0.2pt}}
\paragraph{Discrete Beats Continuous}\aptLtoX{}{\hspace{0.5pt}}%
To evaluate the transform independently of layer selection,
\Cref{tab:sig_transform} compares four transformations using all
layers: count and count$\cdot$idf (Steps~1--2 of
\S\ref{sec:sig_construction}), plus two ablations---
\emph{gate-prob}, the summed softmax gate probabilities over prefill
tokens; and \emph{binary}, the indicator that
an expert appears in the top-$k$ at least once.
The two count-based variants beat the continuous gate-probability baseline by
\textgreater 0.075 in mean~$\rho$. Unlike gate probabilities, discrete
counts record only experts selected during prefill.

\aptLtoX{}{\vspace{-0.5\baselineskip plus -2pt minus -0.2pt}}
\paragraph{{IDF Denoises}}\aptLtoX{}{\hspace{0.5pt}}%
IDF reweighting raises mean~$\rho$ from $0.749$ for plain count to
$0.764$. IDF down-weights common experts, giving rare experts more
influence on signature similarity.
This changes the relative influence of expert counts without changing
the gate's top-$k$ selections.

\begin{figure}[t]
\centering
\includegraphics[width=0.90\columnwidth]{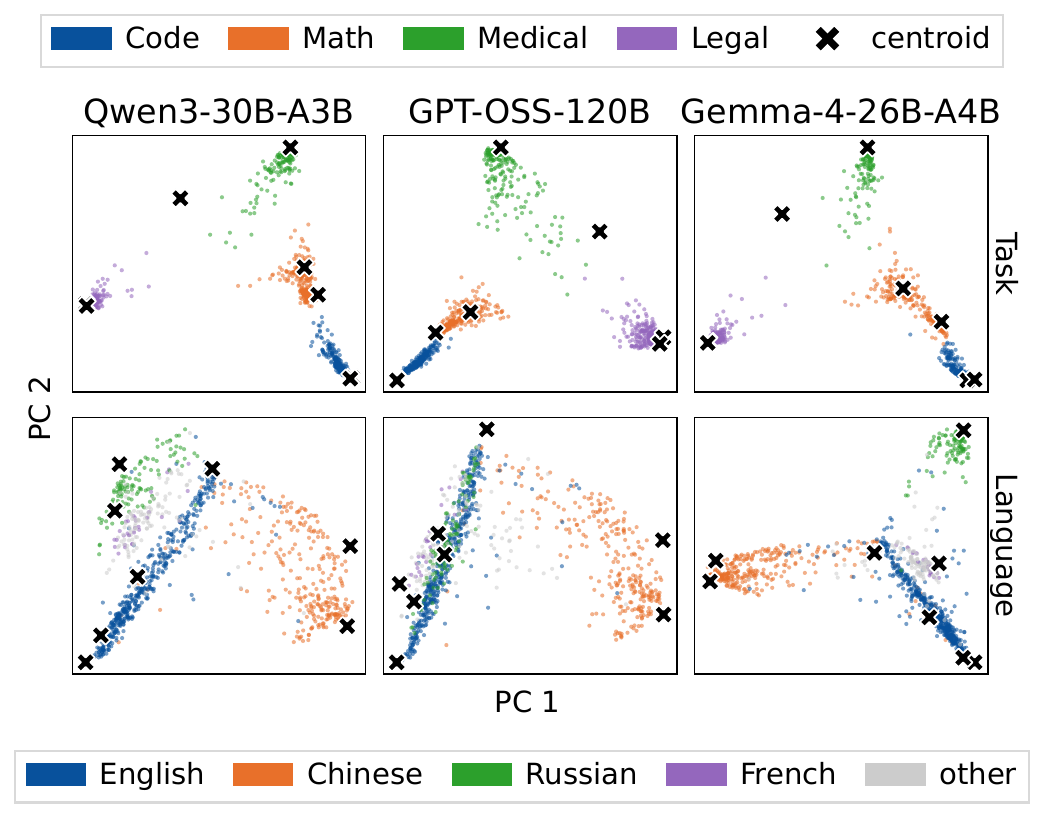}
\vspace{-1em}
\caption{PCA of calibration signatures with Hungarian-balanced centroids
($K{=}8$ for legibility), colored by task domain (top) or top-4
WildChat~\cite{wildchat} language (bottom).}
\Description{Six scatter plots project calibration signatures onto
their first two principal components. Columns show Qwen, GPT-OSS,
and Gemma; rows show Task and Language. Points are colored by domain,
and black crosses mark eight balanced K-means centroids in each
panel. Domains differ in position and spread. In the Qwen panels,
Math signatures form a relatively compact region, while English
signatures span a broader region.}
\label{fig:cluster_pca}
\end{figure}

\aptLtoX{}{\vspace{-0.5\baselineskip plus -2pt minus -0.2pt}}
\paragraph{Layer Mask Compresses}\aptLtoX{}{\hspace{0.5pt}}%
\Cref{fig:sig_layers} plots signature quality~$\rho$ under greedy layer
selection on count$\cdot$idf signatures. Each setting peaks at
$N^{*}<L$, improving over the all-layer $\rho(L)$ by 0.005 to 0.032;
additional layers dilute the signature without improving separation.
\sys's offline fit keeps these $N^{*}$ layers.

\subsection{Decode Clustering and Routing}
\label{sec:design_routing}

Given an expert signature, \sys selects one of $K$ decode workers
based on both expert locality and worker load.
\emph{1) Locality:} requests in the same decode batch should activate
overlapping experts. \emph{2) Load:} non-MoE compute (attention, QKV projection,
and normalization) increases with the number of active requests on a
worker, so excess load can increase tail TPOT
(\cref{sec:challenge_load}).
\sys combines balanced clustering with locality-band routing.
Balanced clustering (\cref{sec:design_clustering}) groups similar signatures
while balancing cluster sizes, with one centroid per decoder.
At the prefill$\,\to\,$decode handoff,
locality-band routing (\cref{sec:design_route}) selects the least-loaded
worker among the candidates in the request's locality band.

\aptLtoX{}{\vspace{-0.25\baselineskip plus -1pt minus -0.1pt}}
\subsubsection{Balanced Clustering}
\label{sec:design_clustering}

\sys uses Hungarian-balanced $K$-means~\cite{balancedkmeans} to produce
one centroid per decode worker, grouping similar signatures for locality
while balancing group sizes across workers. Standard
$K$-means~\cite{kmeans} with cosine distance favors locality but does not
constrain cluster sizes. On skewed data, uneven clusters can concentrate
requests on a subset of decoders, increasing the risk of load imbalance
and tail-latency spikes. At each iteration, the Hungarian
algorithm finds the assignment minimizing total cosine distance for the
current centroids, with at most $\lceil N/K \rceil$
signatures per centroid, where $N$ is the number of input signatures.
\cref{sec:design-validation} evaluates this balance constraint.

\begin{samepage}
\Cref{fig:cluster_pca} shows the result of applying balanced $K$-means to
the calibration signatures, with fitted centroids overlaid on their
two-dimensional PCA projection. The centroids are fitted to group similar
signatures while constraining cluster sizes, promoting both expert
locality and load balance across decode workers. The projection also
reveals differences in within-domain spread, including relatively
compact Math signatures and broadly distributed English signatures.
\end{samepage}

\begin{figure}[t]
\centering
\includegraphics[width=0.95\columnwidth]{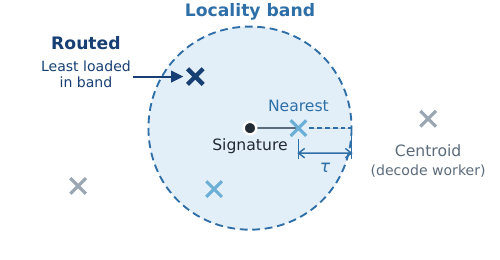}
\vspace{-1em}
\caption{Locality-band routing. Each $\times$ denotes a decoder centroid.
The circle marks candidates within $\tau$ of the best cosine similarity
to the request signature. \sys selects the least-loaded candidate,
not necessarily the nearest.}
\Description{The black dot is a request signature, and each cross is
the centroid of a decode worker. A dashed circle marks the locality band.
Radial distance schematically represents cosine distance, and tau marks
the gap from the nearest centroid to the band boundary. The nearest
centroid lies to the right of the signature. A horizontal arrow identifies
the dark navy centroid in the upper-left part of the band as the routed
decoder, which has the fewest in-flight requests among eligible workers.
Another light-blue centroid is also inside the band. Two gray centroids
outside the band are excluded, regardless of their load.}
\label{fig:locality_band_routing}
\end{figure}

Centroids are initialized from calibration signatures and periodically
refitted using recent request signatures. The IDF weights and layer mask
remain fixed after calibration. To keep each decoder associated with a
similar signature region after refitting, we match the new centroids
one-to-one to the current centroids by maximizing total cosine similarity.
Refitting and
matching run in the background on the CPU, off the request-serving critical
path. Requests continue using the current centroids without waiting for the
update. Once refitting and matching are complete, the router atomically
switches to the updated centroids. If the decode pool size changes,
\sys can rerun the same balanced $K$-means with $K$ set to the new number
of decode workers. \cref{sec:eval_generalization} evaluates centroid
refitting under workload drift.

\aptLtoX{}{\vspace{-0.25\baselineskip plus -1pt minus -0.1pt}}
\subsubsection{Locality-Band Routing}
\label{sec:design_route}

Balanced cluster sizes do not guarantee balanced loads during serving.
Pure top-1 routing---sending each request to its nearest centroid---
ignores current load, so a burst near one centroid can overload its
worker. \sys's online routing rule,
locality-band routing, favors locality when the signature
singles out one worker, and falls back to load-aware balancing when
several workers are nearly tied.

For each request, \sys computes the cosine similarity $\sigma_k$
between its expert signature and each of the $K$ centroids; let
$\sigma^* = \max_k \sigma_k$ denote the highest similarity.
Workers with $\sigma_k \ge \sigma^* - \tau$ form the \emph{locality band}.

As shown in \Cref{fig:locality_band_routing}, \sys selects the worker with
the fewest in-flight decode requests within this band. The router tracks this load as
requests are dispatched and complete. The parameter $\tau \in [0, 1]$
interpolates between pure top-1 routing ($\tau{=}0$) and pure
shortest-queue routing ($\tau{=}1$).

The locality band adapts to signature confidence. When a signature
strongly favors one centroid, the band contains fewer workers,
preserving expert locality. When several centroids match similarly
well, the band includes more workers, providing greater flexibility
to balance load. \sys uses $\tau{=}0.1$ by default.
\cref{sec:design-validation} evaluates this choice.

\subsection{Prefix Cache Coherence}
\label{sec:design_prefix_cache}

\sys preserves the full prompt's expert signature across cache hits by
storing expert-activation counts alongside the KV cache. Because the prefix
cache is managed at KV-block granularity (\cref{sec:background_prefix}),
\sys stores these counts at the same granularity, using the same block IDs.
This lets each reused prefix block supply the expert-activation counts
for its own tokens, rather than relying on a signature for the earlier
request's entire prompt.

\begin{figure}[t]
\centering
\includegraphics[width=0.95\columnwidth]{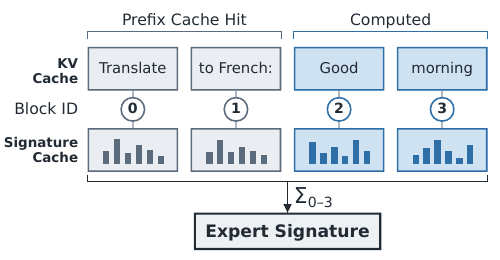}
\vspace{-1em}
\caption{\sys stores expert-activation counts at KV-block granularity,
co-indexed with the KV cache. Cached prefix counts are combined with
counts from newly computed tokens to construct the full prompt's
expert signature.}
\Description{A prompt is divided into four KV-cache blocks with IDs
0 through 3. The first two blocks are prefix-cache hits, and the
remaining two are newly computed. Each KV block has matching
expert-activation counts in the signature cache. Counts from all four
blocks are summed to reconstruct the prompt's expert signature,
combining the cached prefix with newly computed tokens.}
\label{fig:prefix_cache_coherence}
\vspace{-0.5\baselineskip}
\end{figure}

As prefill writes a token's K/V into a block, \sys increments that
block's per-layer counts for the token's top-$k$ experts.
A request's KV cache spans a block set $\mathcal{B}(r)$. Its raw counts
are reconstructed by summing its block rows,
$c_r = \sum_{b \in \mathcal{B}(r)} \mathrm{sig}[b]$. The cache stores
additive counts rather than normalized signatures, so blocks from different
prefills can be summed before constructing a complete request signature.

After summing these counts, the router applies the fixed IDF weights and
layer mask, then L2-normalizes the result to obtain $s_r$. On a partial hit
(\Cref{fig:prefix_cache_coherence}), cached rows supply the prefix counts
and fresh rows supply the remaining counts; their sum matches a cold
prefill of the same prompt. The
signature cache follows the lifecycle of KV-cache blocks: reusing a cached prefix
preserves its counts, while reassigning a block to new tokens resets them.
Thus, \sys keeps no prefix-tree, eviction, or sharing state of its own.

\sys's signature cache is a preallocated GPU tensor with one slot per
KV-cache block, using one int8 per (block, MoE layer, expert); its total
size is bounded by $\text{num\_gpu\_blocks} \cdot L \cdot E$ bytes. For
the three models in the main evaluation with tensor parallelism (TP)
degree 1 ($L \le 48$, $E{=}128$),
each block's row occupies at most $6$\,KiB. Each forward accumulates the
new tokens' expert picks into the matching rows in a single GPU pass.
At prefill completion, block rows are summed into per-request $[L,E]$
count tensors and copied to the host. No separate block allocator or
eviction policy is needed.
\cref{sec:design-validation} evaluates TPOT with prefix caching enabled
and disabled.

\section{Implementation}
\label{sec:implementation}

\sys is implemented on top of vLLM~\cite{vllm} as three components: a
router that dispatches each request to a decoder by signature locality,
engine-side hooks that capture expert-activation counts during prefill, and a
calibration pipeline that clusters captured signatures into per-decoder
centroids---roughly $2{,}000$ lines of new Python.

\aptLtoX{}{\vspace{-0.5\baselineskip plus -2pt minus -0.2pt}}
\paragraph{Signature Path}\aptLtoX{}{\hspace{0.5pt}}%
A prefill hook accumulates per-layer per-expert activation counts at KV-block
granularity, stored in an expert signature cache keyed by block ID. At prefill
completion, the prompt's per-block counts are summed and piggybacked onto
the existing prefill--decode handoff---no new RPC channel is required.

\aptLtoX{}{\vspace{-0.5\baselineskip plus -2pt minus -0.2pt}}
\paragraph{Per-Decoder EP Load Balancing}\aptLtoX{}{\hspace{0.5pt}}%
METRO~\cite{metro} balances expert-parallel ranks by
\emph{activated-expert} load rather than token count. For this objective,
we estimate each cluster's expected per-expert decode activation from the
calibration trace, then for every (decoder, layer) greedily assign the next
most-active expert to the lightest-loaded rank. Since each decoder serves a distinct cluster, a single
global EP layout would balance one cluster at the others' expense; per-decoder
placement matches each decoder to the workload it sees while holding the
per-rank expert count fixed.

\aptLtoX{}{\vspace{-0.5\baselineskip plus -2pt minus -0.2pt}}
\paragraph{Calibration}\aptLtoX{}{\hspace{0.5pt}}%
A single calibration script processes the captured trace and produces the
artifacts loaded at startup. These include a routing artifact containing
the layer mask, IDF weights, and $K$ Hungarian-balanced centroids in the
masked, IDF-weighted, L2-normalized space. For multi-GPU expert parallelism,
the script also produces an expert placement file. During serving, the
layer mask, IDF weights, and expert placement remain fixed; centroids are
periodically refreshed as described in~\cref{sec:design_clustering}.

\aptLtoX{\begin{figure*}[t!]
\centering
\includegraphics[width=0.96\linewidth]{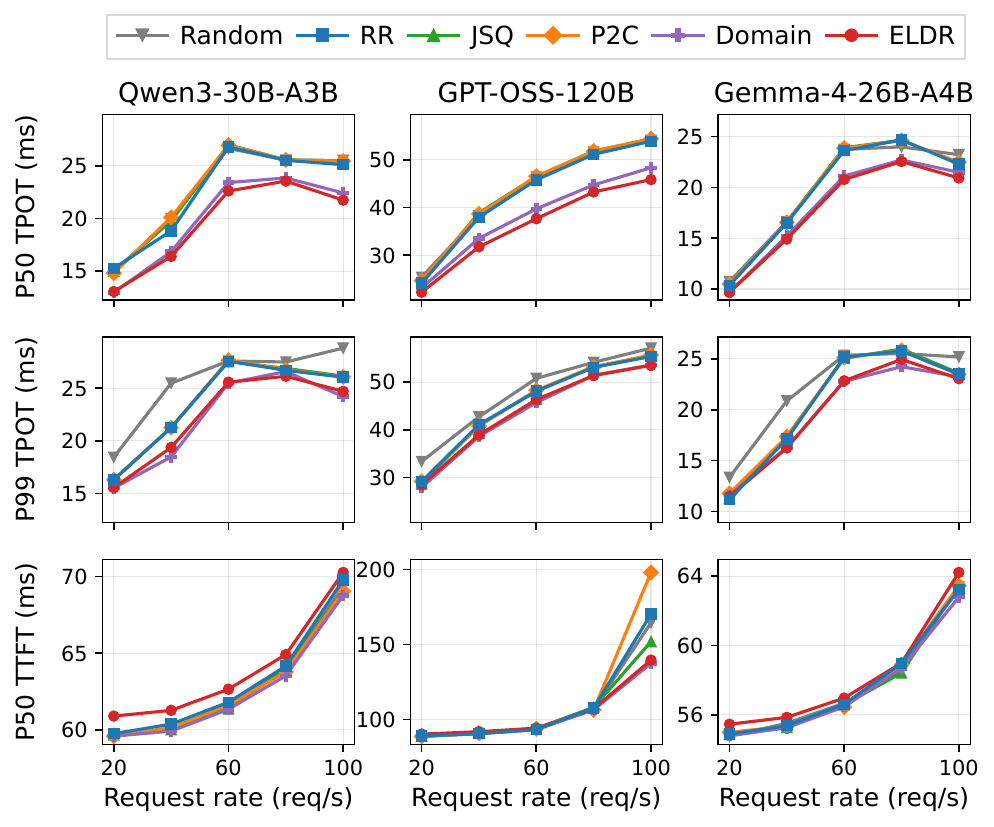}
\caption{Task workload on 8P16D: P50/P99~TPOT and P50~TTFT
across request rates.}
\Description{Nine line charts compare Random, RR, JSQ, P2C, Domain,
and ELDR on Task. Columns show Qwen, GPT-OSS, and Gemma; rows show
P50 TPOT, P99 TPOT, and P50 TTFT. The horizontal axis spans 20 to
100 requests per second, and latency is measured in milliseconds.
ELDR has lower median TPOT than RR at every tested rate. Tail
improvements are smaller, and Domain remains competitive.}
\label{fig:main_task}
\end{figure*}
\begin{figure*}
\centering
\includegraphics[width=0.96\linewidth]{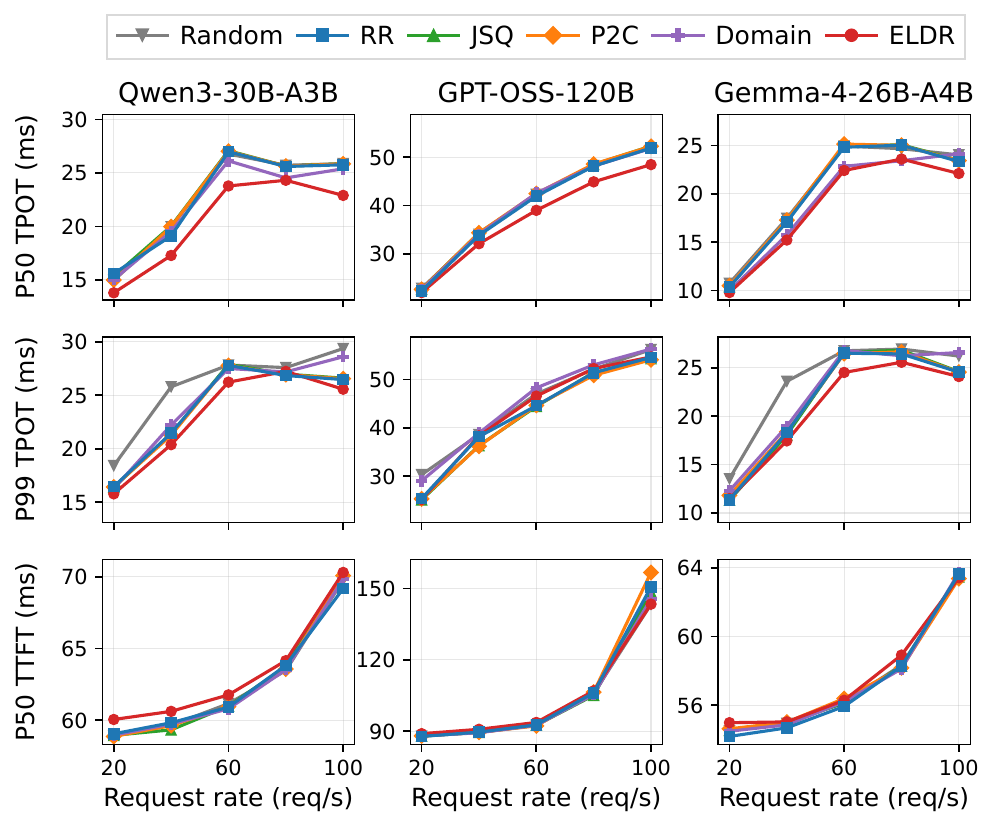}
\caption{Language workload on 8P16D: P50/P99~TPOT and P50~TTFT
across request rates.}
\Description{Nine line charts compare Random, RR, JSQ, P2C, Domain,
and ELDR on Language. Columns show Qwen, GPT-OSS, and Gemma; rows
show P50 TPOT, P99 TPOT, and P50 TTFT across 20 to 100 requests per
second. ELDR lowers median TPOT relative to RR at every tested rate,
but P99 gains are not uniform: GPT-OSS has a small average P99
regression. TTFT curves are largely similar across policies.}
\label{fig:main_language}
\end{figure*}}{\begin{figure*}[t!]
\centering
\begin{minipage}[t]{0.49\textwidth}
\centering
\includegraphics[width=0.96\linewidth]{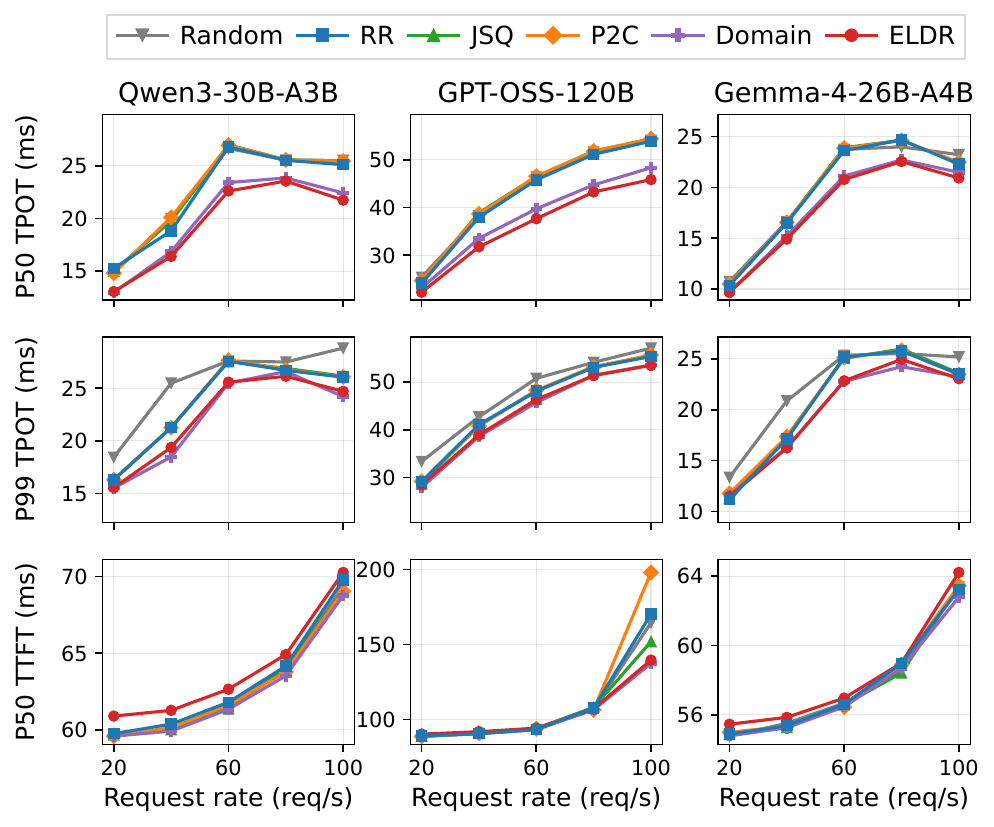}
\caption{Task workload on 8P16D: P50/P99~TPOT and P50~TTFT
across request rates.}
\Description{Nine line charts compare Random, RR, JSQ, P2C, Domain,
and ELDR on Task. Columns show Qwen, GPT-OSS, and Gemma; rows show
P50 TPOT, P99 TPOT, and P50 TTFT. The horizontal axis spans 20 to
100 requests per second, and latency is measured in milliseconds.
ELDR has lower median TPOT than RR at every tested rate. Tail
improvements are smaller, and Domain remains competitive.}
\label{fig:main_task}
\end{minipage}\hfill
\begin{minipage}[t]{0.49\textwidth}
\centering
\includegraphics[width=0.96\linewidth]{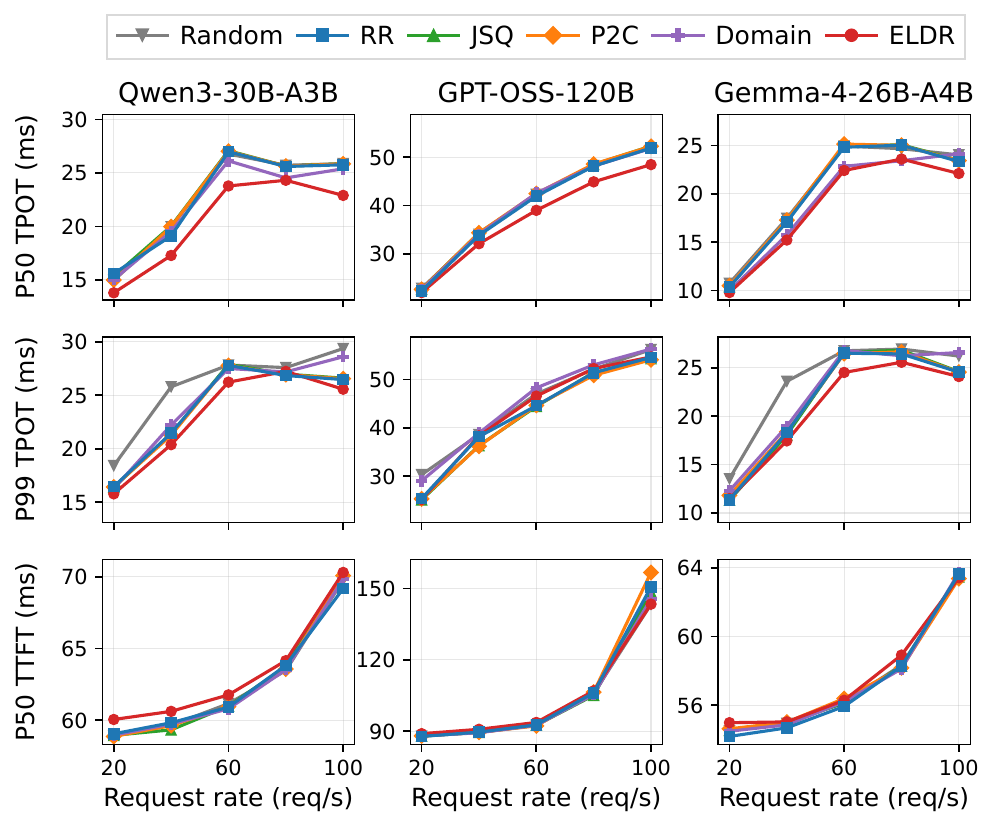}
\caption{Language workload on 8P16D: P50/P99~TPOT and P50~TTFT
across request rates.}
\Description{Nine line charts compare Random, RR, JSQ, P2C, Domain,
and ELDR on Language. Columns show Qwen, GPT-OSS, and Gemma; rows
show P50 TPOT, P99 TPOT, and P50 TTFT across 20 to 100 requests per
second. ELDR lowers median TPOT relative to RR at every tested rate,
but P99 gains are not uniform: GPT-OSS has a small average P99
regression. TTFT curves are largely similar across policies.}
\label{fig:main_language}
\end{minipage}
\end{figure*}}

\begin{samepage}
\section{Evaluation}
\label{sec:evaluation}

\paragraph{Testbed}\aptLtoX{}{\hspace{0.5pt}}%
We evaluate \sys on a five-node cluster, with
eight AMD MI300X GPUs (192~GB HBM each) per node. Each node has eight
ConnectX-7 HCAs, one per GPU, providing 400~Gb/s NDR InfiniBand links.
Our implementation builds on vLLM~\cite{vllm} 0.21.0rc1 with ROCm 7.2
and uses the NIXL~\cite{nixl} connector to transfer KV caches between
prefill and decode workers.

\end{samepage}

\aptLtoX{}{\vspace{-0.5\baselineskip plus -2pt minus -0.2pt}}
\paragraph{Models}\aptLtoX{}{\hspace{0.5pt}}%
We exercise four open MoE models. Three are served at
$TP{=}1$: \textbf{Qwen3-30B-A3B}~\cite{qwen3} (128 experts, top-8, bf16),
\textbf{GPT-OSS-120B}~\cite{gptoss} (128 experts, top-4, mxfp4), and
\textbf{Gemma-4-26B-A4B}~\cite{gemma4} (128 routed experts, top-8, bf16).
For the large-MoE scaling study, \textbf{Qwen3-235B-A22B}~\cite{qwen3}
(128 experts, top-8, bf16) is served with $TP{=}4$ / $EP{=}4$ per instance.

\aptLtoX{}{\vspace{-0.5\baselineskip plus -2pt minus -0.2pt}}
\paragraph{Datasets}\aptLtoX{}{\hspace{0.5pt}}%
We organize requests into two workload families.
\textbf{Task} combines 12 public benchmarks---four code, four math,
three medical, and one legal---into an 11,668-prompt mixture.
Its naturally unequal domain shares reflect source-dataset sizes:
3,600 \emph{legal}~\cite{lexglue},
2,779 \emph{code}~\cite{humaneval, bigcodebench, ds1000, mbpp},
2,744 \emph{math}~\cite{gsm8k,math500,olympiadbench,aquarat},
and 2,545 \emph{medical}~\cite{medqa, pubmedqa, mmlu}\footnote{We use the
professional-medicine subset of MMLU~\cite{mmlu}.} --- a 1.41$\times$
ratio between the largest and smallest domains that stresses load balance when
routing by signature. \textbf{Language} is a 14,000-prompt subset
of WildChat~\cite{wildchat} that exhibits the heavy language skew
shown in~\Cref{fig:wildchat}: English, Chinese, Russian, and
French account for 87.7\% of the volume.
For each workload, the 1,000-prompt
offline calibration split is disjoint from the evaluation
set.

\aptLtoX{}{\vspace{-0.5\baselineskip plus -2pt minus -0.2pt}}
\paragraph{Topologies and Workload}\aptLtoX{}{\hspace{0.5pt}}%
Unless noted, the main evaluation uses an 8-prefiller / 16-decoder
(8P16D) topology, with each instance at $TP{=}1$ so the 24 GPUs
span three nodes. We additionally sweep 8P8D and 8P24D for topology
generalization, and 2P8D~$\times$~($TP{=}4$, $EP{=}4$) for the
235B run, the largest deployment in this study at 40 GPUs spanning
all five nodes of the cluster. Requests follow a Poisson arrival process at
offered rates from 20 to 100~req/s (24--56~req/s for 235B), each rate
held for 120~s after warmup. Warmup requests are excluded from all
reported metrics. We cap each request's output at 512 tokens and enforce
\texttt{ignore\_eos}.

\aptLtoX{}{\vspace{-0.5\baselineskip plus -2pt minus -0.2pt}}
\paragraph{Routers and Baselines}\aptLtoX{}{\hspace{0.5pt}}%
We use separate routing policies for the prefill
and decode phases. For prefill, PrefixHash sends requests sharing a prefix to
the same worker to improve prefix-cache reuse. If that worker's load
exceeds $1.25\times$ the average prefill load, we choose the
least-loaded worker instead. For decode, we compare \sys against four
load-balancing baselines---Random,
RR, JSQ~\cite{jsq}, and P2C~\cite{p2c}---and Domain, an
oracle-label locality baseline. Domain receives each request's task or
language label without online classification and uses JSQ within the
label's preassigned decoder pool. Decoder pool sizes follow the domain proportions
in the calibration set, with
low-frequency domains sharing a pool.
For each combination of model, workload, and request rate, all six policies share the same prefill
routing policy and the same decoder pool, isolating the effect of decode routing.

\aptLtoX{}{\vspace{-0.5\baselineskip plus -2pt minus -0.2pt}}
\paragraph{\sys Configuration}\aptLtoX{}{\hspace{0.5pt}}%
Unless otherwise noted, we use the design
from~\cref{sec:design}: count$\cdot$idf signatures restricted to a
greedy $\rho$-selected layer subset, Hungarian-balanced $K$-means
with $K{=}D$ (one cluster per decoder), and $\tau{=}0.1$
locality-band routing. The 235B run additionally enables per-cluster
per-layer expert-rank permutation for EP load balancing. Signature
capture takes 4--15 minutes per (model, workload) for the 1,000
calibration prompts; the initial fit (greedy mask selection plus
balanced $K$-means) completes in under 10 seconds. The IDF weights and
layer mask remain fixed throughout each serving run. Centroids are initialized
from the calibration signatures and periodically refitted in the background using
a 5-second history of recent request signatures and a 1-second refit interval.
In an instrumented Qwen3-30B-A3B Task run (8P16D, 60~req/s),
background CPU refitting and centroid matching together average
12.90~ms of wall-clock time and 9.99~ms of thread CPU time per update.

\newpage
\subsection{Main Results}
\label{sec:eval_main}
RR achieves the lowest median TPOT among the four load-balancing baselines,
averaged across the five request rates for every model and workload.
\sys reduces median TPOT at every rate across all three models and both
workloads. We report each model's
average and largest per-rate percentage reductions relative to RR.

On Task (Fig. \ref{fig:main_task}), \sys reduces median TPOT by
$8.6$--$14.5\%$ on average and up to $17.8\%$.
P99 TPOT decreases by $3.4$--$5.6\%$ on average and up to
$9.2\%$. Relative to the oracle-label Domain baseline, \sys
reduces median TPOT by $1.4$--$4.6\%$ on average, while Domain
retains slightly lower average P99 TPOT.
\sys thus improves median decode latency over both load-only
and oracle-label routing.

On Language (Fig.~\ref{fig:main_language}), \sys reduces median TPOT by
$5.4$--$9.8\%$ on average and up to $11.9\%$.
P99 TPOT decreases by approximately $3.3\%$ on average for Qwen and
Gemma, but increases by $1.2\%$ for GPT-OSS; the peak P99 reduction
is $7.5\%$. Despite oracle language labels, Domain provides
smaller, less consistent median gains and increases average P99 TPOT
relative to RR for all three models. Relative to Domain, \sys
reduces median and P99 TPOT by $2.9$--$7.9\%$ and $4.6$--$7.0\%$
on average, respectively. Requests within the same language can have
different expert-activation patterns. \sys's request-level signatures
capture differences that language labels miss, providing a finer-grained
basis for locality-aware routing.

\begin{figure}[t!]
\centering
\includegraphics[width=0.95\columnwidth]{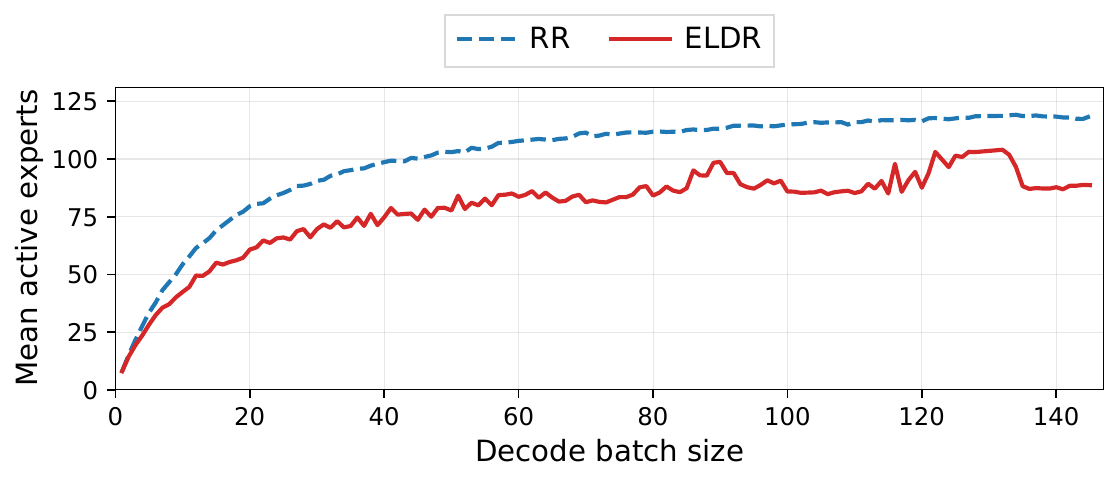}
\caption{Active-expert reduction on Qwen3-30B-A3B (Task, 8P16D):
mean active experts per decode step across batch sizes.
Lower counts indicate greater expert reuse within a batch.}
\Description{A line chart plots mean active experts per decode step
against decode batch size. The dashed blue RR curve rises rapidly
and approaches the full expert pool. The solid red ELDR curve stays
lower at matched batch sizes, indicating greater within-batch expert
reuse. The mean reduction relative to RR is 22 percent.}
\label{fig:expert_reduction}
\end{figure}

To relate these gains to expert reuse, \Cref{fig:expert_reduction}
compares per-step active-expert counts for Qwen3-30B-A3B
(Task, 8P16D) over approximately 50K decode steps. At matched batch sizes,
\sys activates $22.0\%$ fewer experts on average than RR.
Because batch size is held fixed, this reduction reflects greater expert
overlap among requests, rather than fewer requests being processed.
This overlap enables greater reuse of expert weights within each decode
step, consistent with the relationship between active-expert count and
latency established in \cref{sec:motivation_active_experts}.

Averaged across the five rates, median TTFT increases by
$0.4$--$0.8$~ms ($0.7$--$1.3\%$) for Qwen and Gemma. For GPT-OSS,
median TTFT decreases by $3.0\%$ on Task and remains nearly unchanged
on Language. Overall, \sys reduces TPOT with little net impact on
median TTFT. Its lightweight signature path reuses expert selections
computed during prefill. Background centroid refitting does not block
routing with the current centroids.

To assess repeatability, we ran ten paired RR--\sys trials on
GPT-OSS-120B (Language, 8P16D, $60$\,req/s), using identical prompts and
seeded Poisson schedules within each pair. Execution order alternated,
and all pairs were analyzed. \sys reduced median TPOT in every pair,
averaging $5.60\%$ (95\% paired-bootstrap confidence interval:
$4.63$--$6.53\%$).

\subsection{Design Validation}
\label{sec:design-validation}

\begin{figure}[t!]
\centering
\includegraphics[width=0.95\columnwidth]{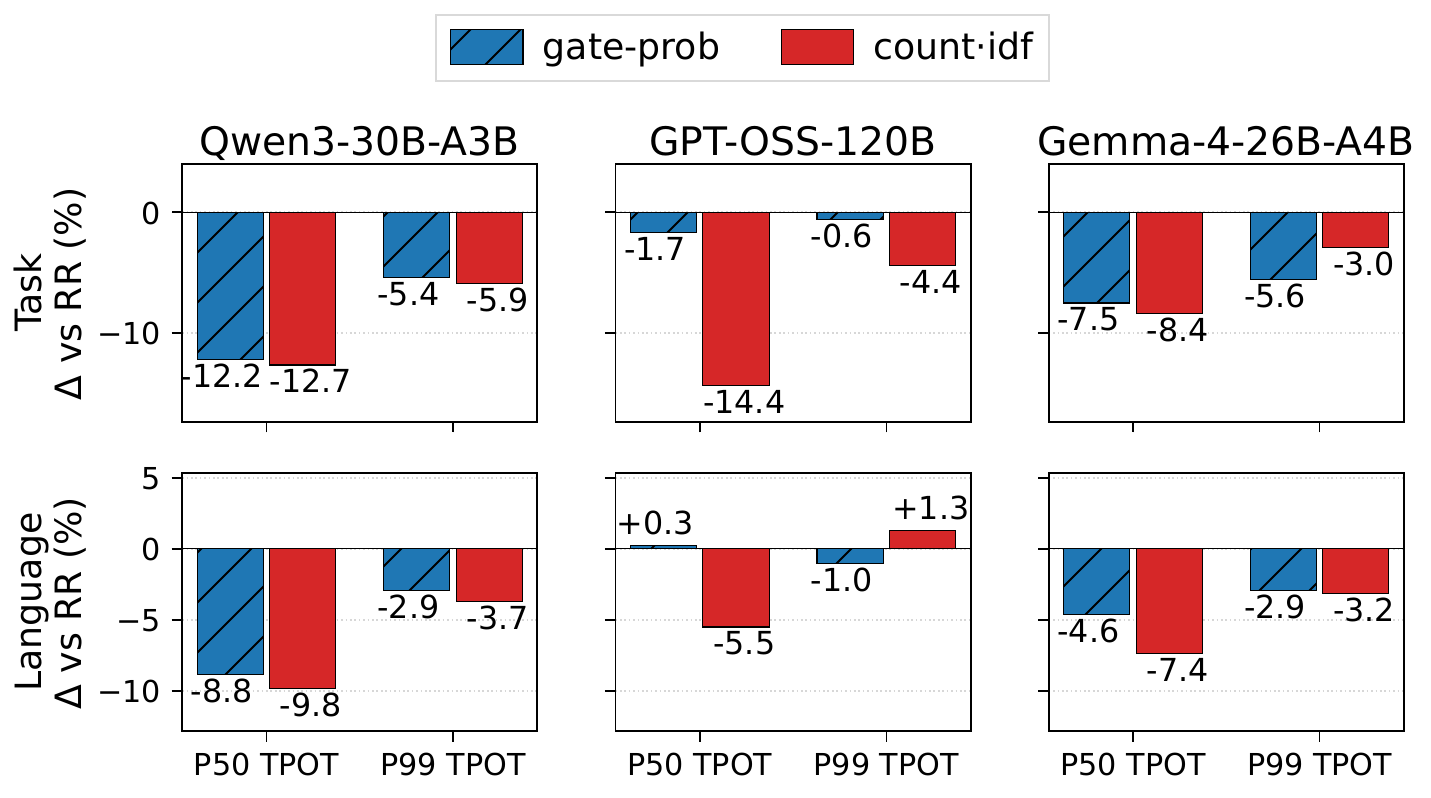}
\caption{Signature ablation on 8P16D: P50/P99 TPOT changes relative to
matched RR controls (\%), averaged across $20$, $40$, $60$, $80$, and $100$\,req/s.
\sys's IDF-reweighted top-$k$ count signature (count$\cdot$idf)
is compared with gate-probability signatures. Lower is better.}
\Description{Six grouped bar charts compare gate-probability signatures
in diagonally hatched blue with count-idf signatures in solid red. Columns show Qwen,
GPT-OSS, and Gemma; rows show Task and Language. Each panel reports
mean P50 and P99 TPOT changes relative to matched RR, with negative
values indicating reductions. Count-idf gives larger P50 reductions
in all six settings. P99 comparisons are mixed, with gate-probability signatures
performing better for Gemma Task and GPT-OSS Language.}
\label{fig:transform_ablation}
\end{figure}

\paragraph{Expert Signature}\aptLtoX{}{\hspace{0.5pt}}%
\Cref{fig:transform_ablation} compares \sys's count$\cdot$idf signature
(\cref{sec:design_signature}) with softmax gate probabilities, the
strongest continuous alternative.
Both use greedy layer selection, balanced $K$-means
($K{=}16$), and $\tau{=}0.1$. Averaged across the five rates,
count$\cdot$idf's median TPOT reductions relative to matched RR exceed
gate-prob's by $0.5$--$12.7$ percentage points across all six
model--workload pairs.
These results support count$\cdot$idf for lower median decode latency.

\begin{figure}[t!]
\centering
\includegraphics[width=0.95\columnwidth]{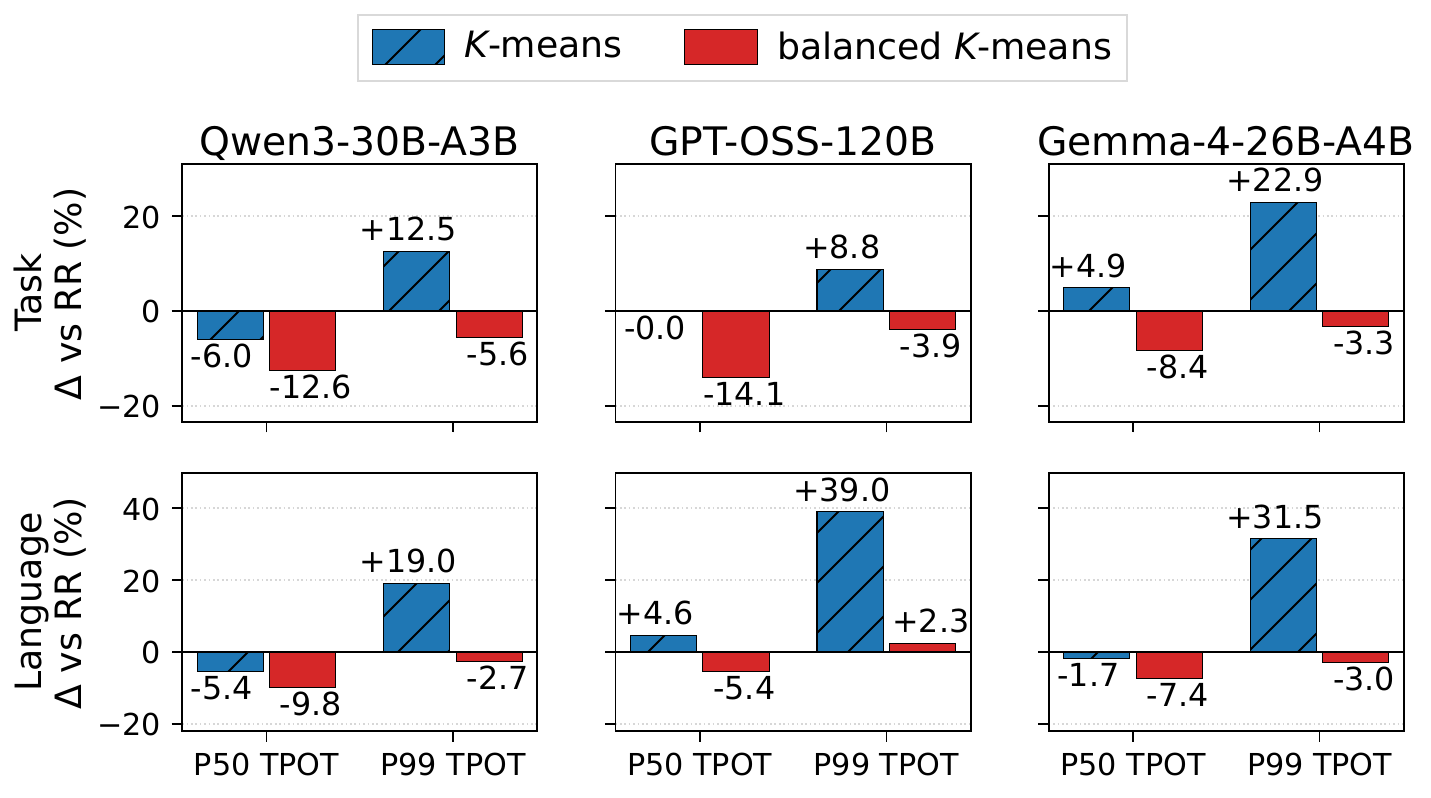}
\caption{Cluster-balance ablation on 8P16D: P50/P99 TPOT changes relative to
matched RR controls (\%), averaged across $20$, $40$, $60$, $80$, and $100$\,req/s.
\sys's balanced $K$-means is compared with vanilla $K$-means.
Lower is better.}
\Description{Six grouped bar charts compare vanilla K-means in diagonally hatched blue
with balanced K-means in solid red across three models and two workloads.
Each panel reports mean P50 and P99 TPOT changes relative to matched
RR. Balanced K-means improves both metrics relative to vanilla
K-means in every setting. Vanilla K-means increases P99 by 8.8 to
39 percent. Balanced K-means reduces P99 in five settings, with
a 2.3 percent increase for GPT-OSS Language.}
\label{fig:cluster_ablation}
\end{figure}

\aptLtoX{}{\vspace{-0.5\baselineskip plus -2pt minus -0.2pt}}
\paragraph{Cluster Balance}\aptLtoX{}{\hspace{0.5pt}}%
\label{sec:eval_cluster_ablation}
\Cref{fig:cluster_ablation} compares balanced and vanilla $K$-means for
initial fitting and online refitting. Both use $\tau{=}0.1$ and a
$5$-second history, refitting every $1$ second.
Averaged across five rates, balanced $K$-means reduces median TPOT by
$5.4$--$14.1\%$ across six model--workload pairs; vanilla ranges
from a $6.0\%$ reduction to a $4.9\%$ increase.
Vanilla increases P99 TPOT by $8.8$--$39.0\%$ relative to RR.
Balanced $K$-means reduces it by $2.7$--$5.6\%$ in five pairs but
increases it by $2.3\%$ for GPT-OSS Language. Balanced $K$-means thus
improves median and P99 TPOT over vanilla in all six pairs.

\begin{figure}[t!]
\centering
\includegraphics[width=0.95\columnwidth]{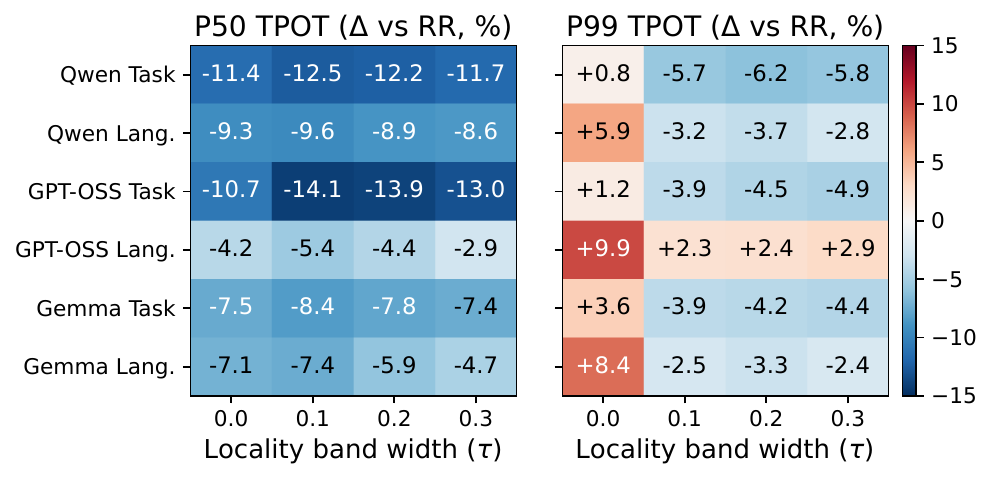}
\caption{Locality-band ablation on 8P16D: P50/P99 TPOT changes relative to
matched RR controls (\%), averaged across $20$, $40$, $60$, $80$, and $100$\,req/s.
\sys's default band width, $\tau{=}0.1$, is compared with
$\tau{=}0$, $0.2$, and $0.3$. Lower is better.}
\Description{Two annotated heatmaps show mean P50 and P99 TPOT changes
relative to matched RR. Rows are six model-workload pairs; columns
are locality-band widths 0, 0.1, 0.2, and 0.3. Blue indicates a
reduction and red an increase. Width 0.1 gives the largest median
reduction in every row. Width 0 increases P99 in every row; wider
bands sometimes reduce P99 further than 0.1. GPT-OSS Language has
a P99 increase at every tested width.}
\label{fig:tau_ablation}
\end{figure}

\aptLtoX{}{\vspace{-0.5\baselineskip plus -2pt minus -0.2pt}}
\paragraph{Locality Band Width}\aptLtoX{}{\hspace{0.5pt}}%
\label{sec:eval_tau_ablation}
\Cref{fig:tau_ablation} compares \sys's default locality-band width,
$\tau{=}0.1$, with $\tau{=}0$, $0.2$, and $0.3$. Averaged across the five request rates,
top-1 routing ($\tau{=}0$) increases P99 TPOT by $0.8$--$9.9\%$
relative to RR in all six model--workload pairs. The default band gives the largest
median TPOT reduction in each model--workload pair ($5.4$--$14.1\%$) and reduces P99 TPOT by
$2.5$--$5.7\%$ in five pairs; GPT-OSS Language still increases
P99 by $2.3\%$. Increasing $\tau$ admits less similar centroids, giving
the router more options to balance runtime load but potentially weakening
expert locality within each batch. Wider bands sometimes reduce P99 further but
yield smaller median gains, supporting $\tau{=}0.1$ as a compromise
between expert locality and runtime load.

\begin{figure}[t!]
\centering
\includegraphics[width=0.95\columnwidth]{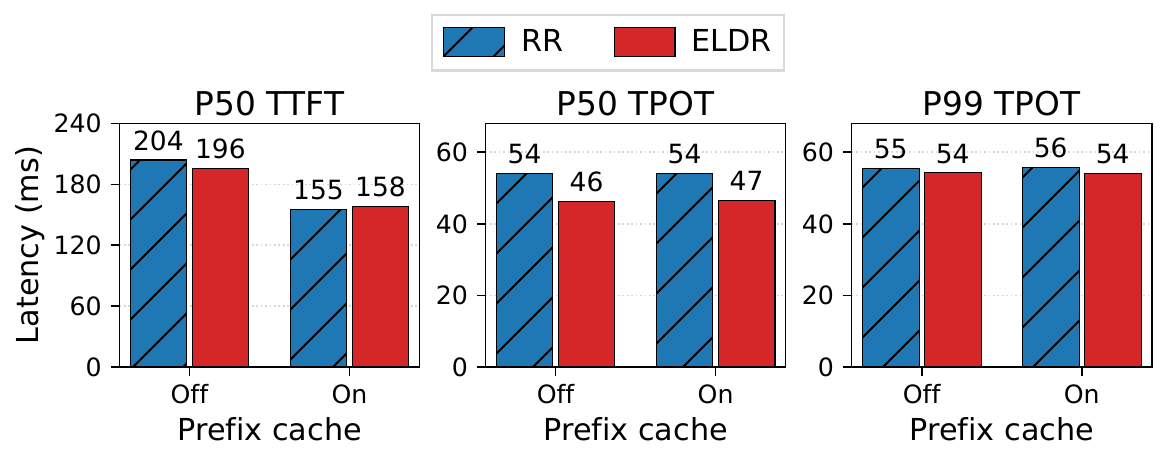}
\caption{Prefix-cache ablation on GPT-OSS-120B (Task, 1P16D, 100\,req/s):
median TTFT and P50/P99 TPOT with caching off and on, using 12,000
requests per condition from a pool of 2,000 prompts.}
\Description{Three grouped bar charts compare RR in diagonally hatched blue with ELDR
in solid red, with prefix caching off and on. Enabling caching lowers
median TTFT from about 204 to 155 milliseconds for RR and from
196 to 158 milliseconds for ELDR. P50 and P99 TPOT change little
between cache states, while ELDR remains below RR in both states.}
\label{fig:prefix_ablation}
\end{figure}

\aptLtoX{}{\vspace{-0.5\baselineskip plus -2pt minus -0.2pt}}
\paragraph{Prefix Cache Composition}\aptLtoX{}{\hspace{0.5pt}}%
\label{sec:eval_prefix_cache}
\Cref{fig:prefix_ablation} shows that \sys's TPOT benefit is preserved
with prefix caching both enabled and disabled. We use a single prefill
worker to concentrate prefix reuse, while retaining 16 decode workers.
Across the two prefix-cache configurations, \sys reduces median TPOT by 13.8--14.3\%
and P99 TPOT by 2.2--2.7\% relative to RR. Enabling the cache additionally
reduces median TTFT by 23.9\% for RR and 19.0\% for \sys,
while leaving TPOT nearly unchanged for both routing policies. These results show that \sys's
TPOT gains complement the reductions in TTFT obtained through prefix caching.

\subsection{Generalization}
\label{sec:eval_generalization}

\paragraph{Single-Domain Workloads}\aptLtoX{}{\hspace{0.5pt}}%
\sys can still exploit expert locality in a single-domain workload
as long as requests exhibit within-domain signature diversity.
\Cref{tab:homogeneous} compares RR and \sys on Math-only and English-only
workloads using Qwen3-30B-A3B at 8P16D across 20--100~req/s. We initialize
centroids from separate Math-only and English-only calibration requests
while keeping signature construction unchanged. Averaged across the five request rates,
\sys reduces median TPOT by $1.5\%$ and $8.3\%$, respectively.
P99 reductions average $0.7\%$ and $0.3\%$.
The larger median gain on English-only is consistent with the Qwen panels
in~\Cref{fig:cluster_pca}: Math signatures are relatively concentrated,
whereas English signatures span multiple regions. This within-domain
variation provides more opportunities for \sys to group requests with
similar expert usage.
Workloads with tightly concentrated request signatures, however,
offer limited opportunity for locality-aware routing.

\begin{table}[t!]
\centering
\caption{TPOT reduction relative to RR on single-domain workloads
(Qwen3-30B-A3B, 8P16D), averaged across $20$, $40$, $60$, $80$, and $100$\,req/s.}
\label{tab:homogeneous}
\begin{tabular}{lrr}
\toprule
Workload & P50 TPOT & P99 TPOT \\
\midrule
Math-only & $1.5\%$ & $0.7\%$ \\
English-only & $8.3\%$ & $0.3\%$ \\
\bottomrule
\end{tabular}
\end{table}

\begin{figure}[t!]
\centering
\includegraphics[width=0.95\columnwidth]{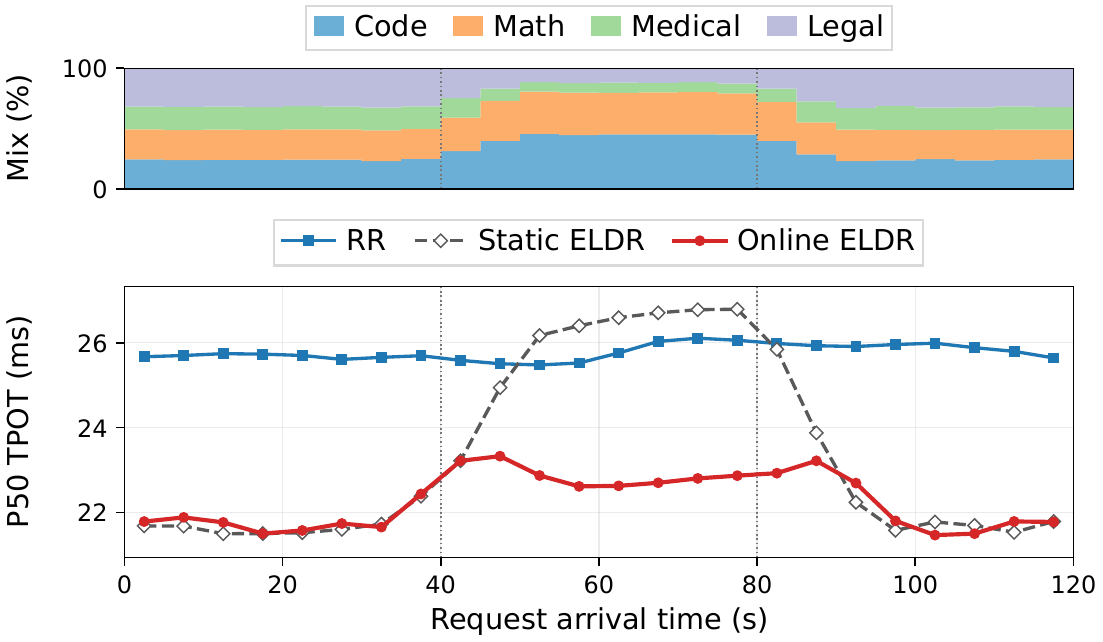}
\caption{Workload drift on Qwen3-30B-A3B (8P16D, 100~req/s).
Top: Task-domain mix in the RR run. Bottom: median whole-request TPOT in 5-second
request-arrival bins. Online \sys uses a 5-second history and refits
centroids in the background every second.}
\Description{Two aligned plots share a request-arrival time axis
from 0 to 120 seconds. The upper stacked-area plot shows the
Code, Math, Medical, and Legal shares, with Code and Math increasing
before the initial mix is restored. The lower plot compares median
whole-request TPOT in five-second arrival bins for RR, static ELDR,
and online ELDR. During steady drift from 50 to 80 seconds,
online ELDR stays below both static ELDR and RR.}
\label{fig:workload_drift}
\end{figure}

\aptLtoX{}{\vspace{-0.5\baselineskip plus -2pt minus -0.2pt}}
\paragraph{Workload Drift}\aptLtoX{}{\hspace{0.5pt}}%
\Cref{fig:workload_drift} evaluates \sys under a changing request mix on
Qwen3-30B-A3B at 8P16D and 100~req/s. We increase the combined code/math
share of the Task workload from 49\% to 80\% and then restore the initial
mix. We compare RR, static \sys with fixed centroids, and \sys's default
configuration, which periodically refits centroids in the background.
Both \sys variants share initial centroids
and $\tau{=}0.1$. The online variant uses a 5-second history and a
1-second refit interval. During steady drift (50--80~s), online \sys's
median TPOT, averaged across the 5-second bins, is 14.4\% lower than
static \sys's and 11.9\% lower than RR's.
Thus, centroid refresh preserves locality benefits as the
request mix changes.

\begin{table}[t!]
\centering
\caption{Topology generalization on Qwen3-30B-A3B (Language):
P50/P99 TPOT changes relative to RR (\%), averaged across
$20$, $40$, $60$, $80$, and $100$\,req/s.}
\label{tab:topology}
\begin{tabular}{lrrr}
\toprule
 & 8P8D & 8P16D & 8P24D \\
\midrule
P50 TPOT & $-8.0\%$ & $-9.8\%$ & $-10.2\%$ \\
P99 TPOT & $-2.5\%$ & $-0.8\%$ & $-1.1\%$ \\
\bottomrule
\end{tabular}
\end{table}

\aptLtoX{}{\vspace{-0.5\baselineskip plus -2pt minus -0.2pt}}
\paragraph{Decoder Pool Size}\aptLtoX{}{\hspace{0.5pt}}%
\Cref{tab:topology} compares 8P8D, 8P16D, and 8P24D on
Qwen3-30B-A3B (Language, $TP{=}1$). Averaged across 20--100~req/s,
\sys reduces median TPOT relative to RR by $8.0\%$, $9.8\%$, and
$10.2\%$, respectively; P99 reductions are $2.5\%$, $0.8\%$, and
$1.1\%$. Median TPOT reductions grow as the decoder pool increases
from 8 to 24 workers, while P99 TPOT improves across all three configurations.

\begin{figure}[t!]
\centering
\includegraphics[width=0.95\columnwidth]{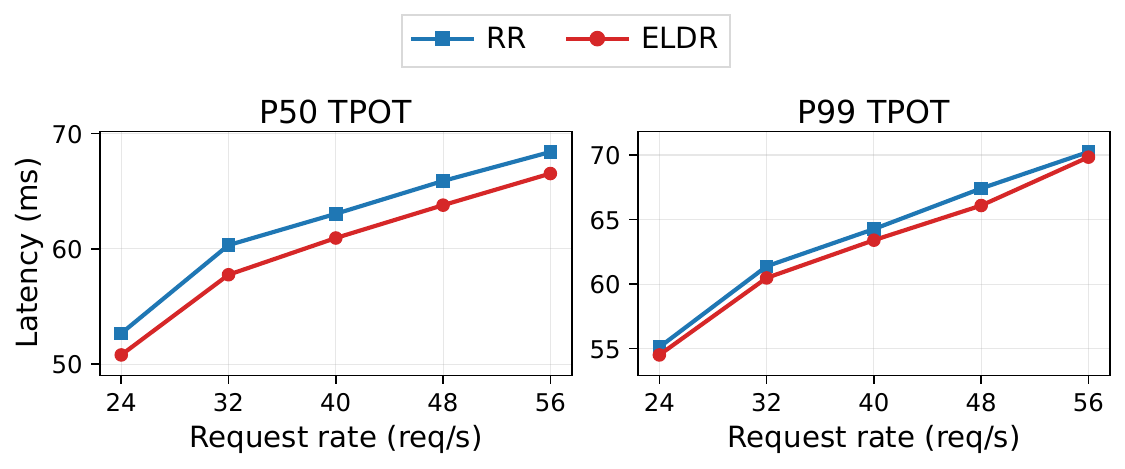}
\caption{Large-model evaluation on Qwen3-235B-A22B (Language, 2P8D,
$TP{=}4$, $EP{=}4$; 40 GPUs across 5 nodes): P50/P99 TPOT across
request rates.}
\Description{Two line charts compare RR in blue with ELDR in red
on Qwen3-235B-A22B. The horizontal axis spans 24 to 56 requests per
second; the vertical axes show P50 and P99 TPOT in milliseconds.
Latency increases with request rate. ELDR remains below RR at every
tested rate in both panels, with a larger relative improvement
in P50 than P99.}
\label{fig:qwen235}
\end{figure}

\aptLtoX{}{\vspace{-0.5\baselineskip plus -2pt minus -0.2pt}}
\paragraph{Large MoE with Expert Parallelism}\aptLtoX{}{\hspace{0.5pt}}%
\Cref{fig:qwen235} evaluates Qwen3-235B-A22B at 2P8D with $TP{=}4$ and $EP{=}4$
per instance (40 GPUs across five nodes). Per-decoder expert placement
(\cref{sec:implementation}) balances cluster-specific expert demand across
the four EP ranks. On Language at 24--56~req/s, \sys reduces median TPOT by
$2.7$--$4.3\%$ and P99 TPOT by $0.6$--$2.0\%$ relative to RR.
Both metrics improve at every tested rate, extending \sys's benefits
to large MoEs with EP.

\section{Related Work}
\label{sec:related}

\paragraph{PD-Disaggregated Serving}\aptLtoX{}{\hspace{0.5pt}}%
PD-disaggregated systems such as DistServe~\cite{distserve},
Splitwise~\cite{splitwise}, and Mooncake~\cite{mooncake} run compute-bound
prefill and memory-bandwidth-bound decode on separate workers, removing
interference and meeting phase SLOs; we adopt this \emph{xPyD} topology.
MemServe~\cite{memserve} integrates prefix caching with disaggregated inference
using pooled GPU and host memory.
OasisKV~\cite{oasiskv} reduces the decode-worker KV-cache footprint and
transfer costs through sparse attention and selective remote prefetching.
Cache-aware schedulers, including SGLang~\cite{sglang} and
Preble~\cite{preble}, exploit prefix-cache locality. Preble jointly considers
prefix-cache reuse and load balance. Neither uses expert overlap as a routing
signal. STAR~\cite{star} addresses evolving decode load by predicting
remaining generation lengths and migrating requests between decode workers.
\sys instead makes placement decisions at the prefill--decode handoff,
considering both predicted expert overlap and current worker load.
Its \emph{expert-activation} locality complements prefix-cache-aware
scheduling without changing the model's outputs.

\aptLtoX{}{\vspace{-0.5\baselineskip plus -2pt minus -0.2pt}}
\paragraph{Expert Parallelism and Locality}\aptLtoX{}{\hspace{0.5pt}}%
Expert parallelism shards experts across devices, where skewed token
routing causes load imbalance across GPUs.
One line of work addresses this imbalance \emph{within} a
deployment by placing or replicating experts and routing tokens across EP
ranks~\cite{eplb,moetuner,leastloadedep}. METRO~\cite{metro} targets
memory-bound decode by reducing the maximum number of activated expert
replicas across EP ranks rather than balancing token counts. For each token,
it selects physical replicas of the gate-selected experts while preserving the logical
token-to-expert assignments. \sys instead places whole requests across
decode workers, grouping requests with overlapping expert usage to reduce
each worker's active-expert union.
Closely related systems exploit expert-activation locality:
Bambhaniya et al.~\cite{eapatterns} group requests using prefill expert activations
and adjust expert placement, while Sem-MoE~\cite{semanticparallelism}
coordinates expert placement with request and token scheduling to colocate
tokens with their experts. These systems primarily target the cost of inter-device
all-to-all \emph{communication} by bringing tokens closer to their experts.
\sys instead routes requests with overlapping expert usage to the same
decode worker, promoting expert-weight reuse within decode batches.
This reduces HBM weight traffic even on single-GPU decode workers.

\aptLtoX{}{\vspace{-0.5\baselineskip plus -2pt minus -0.2pt}}
\paragraph{Batch-Aware Expert Selection}\aptLtoX{}{\hspace{0.5pt}}%
Another line of work addresses the memory-bandwidth bottleneck during decode through approximate
\emph{intra}-worker expert execution.
Lynx~\cite{lynx} remaps low-affinity token-to-expert assignments
to reduce the number of distinct experts invoked. Other approaches
piggyback on already-loaded experts~\cite{opportunistic}, share experts
among the tokens processed together in a batch (XShare~\cite{xshare}), or reroute secondary expert
assignments by similarity (SERE~\cite{sere}).

Unlike these approximate \emph{intra}-worker optimizations, \sys performs
\emph{lossless inter-worker request routing}: it changes request assignment
across decode workers to reduce active-expert sets within decode batches, without changing
token-to-expert assignments.
These approaches are potentially complementary because their decisions operate at
different levels of the serving system: \sys chooses which decode worker serves each request,
while intra-worker methods optimize expert execution within the selected
worker. We therefore compare routing and load-balancing policies while
keeping the model's expert execution unchanged under every policy.

\aptLtoX{}{\vspace{-0.5\baselineskip plus -2pt minus -0.2pt}}
\paragraph{Expert Offloading and Caching}\aptLtoX{}{\hspace{0.5pt}}%
MoE inference systems exploit sparse expert activation to serve
models whose weights exceed GPU memory capacity.
MoE-Infinity~\cite{moeinfinity} uses request-level activation traces to
guide expert caching and prefetching. Pre-gated MoE~\cite{pregatedmoe}
uses trained pre-gates to select the next MoE block's experts in advance,
overlapping weight transfers with computation, while Fiddler~\cite{fiddler} coordinates CPU and GPU
execution to reduce transfer overhead. These systems optimize where experts
are stored and executed when GPU memory capacity is limited. \sys instead targets
GPU-resident PD-disaggregated serving. Keeping all expert weights resident
avoids repeated host-to-device weight transfers, but does not eliminate
accesses to those weights in HBM during each decode step. \sys targets
this per-step bandwidth cost rather than GPU memory capacity. It routes requests with
similar expert signatures to the same decode worker to reduce the number
of distinct experts accessed per decode step, without changing
token-to-expert assignments.
\sys's signature cache has a different purpose from an expert-weight
cache: it stores activation counts, not model parameters, to recover
the routing signal when prefix caching skips part of prefill. It does
not change which expert weights reside in GPU memory.

\section{Conclusion}
\label{sec:conclusion}
Equal load does not mean equal latency in PD-disaggregated MoE serving.
\sys makes \emph{expert locality} a second routing axis through lossless
decode routing. Prefill-derived expert
signatures identify requests likely to share experts during generation;
balanced clustering and locality-band routing reconcile this locality
with live load. Background centroid refitting adapts to workload
changes, while a block-granular signature cache keeps signatures
coherent under prefix caching. Implemented in vLLM, \sys reduces
median TPOT across models, workloads, and topologies. These gains come from
expert-weight reuse within decode batches, not from skipping gate-selected
expert computation. Only request placement changes; model weights and
token-to-expert assignments remain unchanged.

\begin{acks}
We thank the anonymous reviewers and our shepherd for their constructive feedback and guidance.
This research was supported by the MSIT (Ministry of Science, ICT), Korea, under the Global Research Support Program in the Digital Field program (RS-2024-00436680) supervised by the IITP (Institute for Information \& Communications Technology Planning \& Evaluation).
This project is partially supported by Microsoft Research Asia.
This work was supported by the InnoCORE program of the Ministry of Science and ICT (AI Meta-Scientist, N10260110).
\end{acks}

\bibliographystyle{ACM-Reference-Format}
\bibliography{sample-base}%

\end{document}